\documentclass[lettersize,journal]{IEEEtran}
\usepackage{amsmath,amsfonts}
\usepackage{algorithmic}
\usepackage{array}
\usepackage{textcomp}
\usepackage{stfloats}
\usepackage{url}
\usepackage{verbatim}
\usepackage{gensymb}
\usepackage{graphicx}
\graphicspath{{figs/}}
\DeclareGraphicsExtensions{.pdf,.jpeg,.png}
\usepackage{cite}
\usepackage{cleveref}
\usepackage{mdwmath}
\usepackage{mdwtab}
\let\labelindent\relax
\usepackage{enumitem}
\usepackage{eqparbox}
\usepackage{pifont}
\usepackage{subfigure}
\usepackage{float}
\usepackage{multirow}
\usepackage{booktabs}
\usepackage{makecell}
\usepackage{bm}
\usepackage[flushleft]{threeparttable}
\usepackage[linesnumbered,algoruled,boxed,lined]{algorithm2e}
\usepackage[numbers,sort&compress]{natbib}
\usepackage{url}
\usepackage{color}

\fontsize{7.0pt}{\baselineskip}\selectfont
\newcommand{\bd}{{\sc Tuner}\xspace}
\newcommand{\cmark}{\ding{51}}%
\newcommand{\xmark}{\ding{55}}%
\newcommand{\blue}[1]{\textcolor{black}{#1}}

\makeatletter 
\g@addto@macro{\@algocf@init}{\SetKwInOut{Parameter}{Parameters}} 
\makeatother

\begin{document}

\title{Enrollment-stage Backdoor Attacks on Speaker Recognition Systems via Adversarial Ultrasound}

\author{Xinfeng Li, Junning Ze, Chen Yan, Yushi Cheng, Xiaoyu Ji, Wenyuan Xu
\thanks{X. Li, J. Ze, C. Yan (the corresponding author), Y. Cheng, X. Ji, and W. Xu are with USSLAB, Zhejiang University, Hangzhou 310058, China. (Email: xinfengli@zju.edu.cn, zjning@zju.edu.cn, yanchen@zju.edu.cn, yushicheng@zju.edu.cn, xji@zju.edu.cn, wyxu@zju.edu.cn.)}
}




\maketitle


\begin{abstract}
Automatic Speaker Recognition Systems (SRSs) have been widely used in voice applications for personal identification and access control. A typical SRS consists of three stages, i.e., training, enrollment, and recognition. Previous work has revealed that SRSs can be bypassed by backdoor attacks at the training stage or by adversarial example attacks at the recognition stage. 
In this paper, we propose \bd, a new type of backdoor attack against the enrollment stage of SRS via adversarial ultrasound modulation, which is inaudible, synchronization-free, content-independent, and black-box. 
Our key idea is to first inject the backdoor into the SRS with modulated ultrasound when a legitimate user initiates the enrollment, and afterward, the polluted SRS will grant access to both the legitimate user and the adversary with high confidence. 
Our attack faces a major challenge of unpredictable user articulation at the enrollment stage. To overcome this challenge, we generate the ultrasonic backdoor by augmenting the optimization process with random speech content, vocalizing time, and volume of the user.
Furthermore, to achieve real-world robustness, we improve the ultrasonic signal over traditional methods using sparse frequency points, pre-compensation, and single-sideband (SSB) modulation.
\blue{We extensively evaluate \bd on two common datasets and seven representative SRS models, as well as its robustness against seven kinds of defenses. Results show that our attack can successfully bypass speaker recognition systems while remaining effective to various speakers, speech content, etc. To mitigate this newly discovered threat, we also provide discussions on potential countermeasures, limitations, and future works of this new threat.}

\end{abstract}


\section{Introduction}

Automatic speaker recognition systems (SRSs) can identify and authenticate human users by their voices. These systems have been increasingly used in voice assistants and online banking for ensuring that access to critical service and information is only granted to legitimate users~\cite{voice_banking}.
Despite their convenience, SRSs are vulnerable to various types of attacks.
For example, voice spoofing attacks use recorded or synthesized voice samples of legitimate users to deceive the system~\cite{wang2020differences,blue2018hello}. 
Recently, other works have revealed that SRSs are also vulnerable to backdoor attacks~\cite{zhai2021backdoor,gu2017badnets,shi2022audio,luo2022practical} and adversarial example (AE) attacks~\cite{chen2021real,li2020advpulse,li2020practical}. 
However, most backdoor attacks rely on a challenging prerequisite of accessing the training data of the target system; spoofing attacks and AE attacks require generating human-audible sounds, which may be noticed by the victim users when they are nearby, and some attacks even require white-box knowledge that is not available for commercial systems.



\begin{figure}[t]
	\centering
	\includegraphics[width=0.485\textwidth]{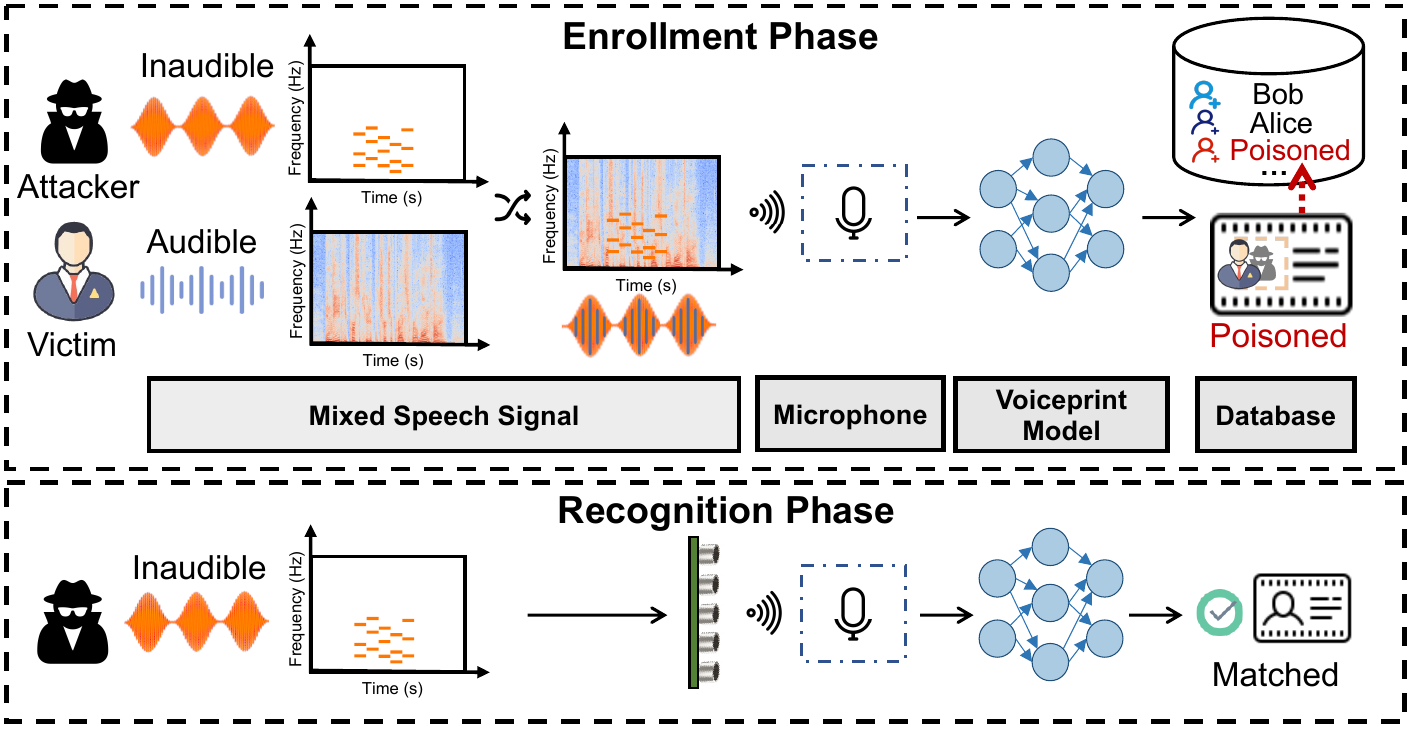}
	\caption{\label{fig:illustration}Illustration of \bd. An adversary emits backdoors modulated on ultrasound when a victim initiates the voiceprint enrollment. Such poisoned voiceprint recorded by the speaker recognition system enables both the adversary and the legitimate user to pass subsequent recognition with relatively high confidence. Particularly, the adversary leverages the inaudible adversarial ultrasound to conduct the backdoor attack in a human-imperceptible manner.}
	\vspace{-10pt}
\end{figure}

In this paper, we propose a backdoor attack named \bd\footnote{We term it ``\bd'' because the backdoor pattern is like a combination of different tones carefully designed by a tuner. \textcolor{black}{Our demo page: https://letterligo.github.io/Tuner/.}} that can bypass commercial speaker recognition systems without accessing the training data, requiring white-box system knowledge, or alarming the victim user.
Different from most of the existing backdoor attacks that target the training stage of SRS models, this paper focuses on the feasibility of backdoor attacks in the enrollment stage, i.e., when legitimate users register their voices for the first time.
As shown in Fig.~\ref{fig:illustration}, the adversary attempts to inject a backdoor into the victim's voice during the user enrollment and poison the voiceprint stored in the speaker database. 
Afterward, the contaminated voiceprints can enable both the legitimate user and the adversary to pass the speaker recognition with high confidence. 

To achieve such an attack, the first requirement we need to meet is to avoid raising user awareness when injecting the backdoor into the voiceprint enrollment, because the victim user will be using the SRS at the same time and can easily notice suspicious sounds from the attacker.
We manage to achieve a completely human-inaudible backdoor injection based on the means of inaudible voice attacks~\cite{zhang2017dolphinattack}, which can make microphones receive audible sounds by emitting inaudible ultrasounds that are beyond the human auditory range (20Hz$\sim$20kHz).
Nevertheless, our investigation reveals that injecting an effective voiceprint backdoor using ultrasound is non-trivial because the backdoor signal will be significantly distorted by the nonlinear distortion and hardware instability during the transmission process.
To cope with the signal distortion, we adapt the backdoor design strategy with ultrasound characteristics and optimize the backdoor signals as a form of combined simple tones at sparse frequency points, as shown in Fig.~\ref{fig:illustration}.
In specific, we first initialize \bd with a modest number of frequencies and select the most effective ones among them. It is worth noting that the selection (i.e., sparse frequency coding) is automatically achieved using the L1-norm regularization term~\cite{lee2006efficient} during the whole optimization process.
This design strategy can better mitigate signal distortion during the ultrasound transmission than existing audible backdoor triggers~\cite{zeng2021rethinking} that have a more complex spectrum similar to human voices. 
We also employ pre-compensation and single-sideband modulation to mitigate signal distortion and boost the transmission efficiency in backdoor delivery.

In addition to the signal distortion, materializing the backdoor attack in real-world scenarios faces a few more challenges. 
(1) During a real-world enrollment, the utterances that the victim user will vocalize are unpredictable, thus the backdoor needs to be content-agnostic.
(2) The voiceprint models deployed in commercial SRSs are generally black-box, i.e., the adversary has no knowledge of the model parameters or the gradient information. 
(3) The attack configuration needs to be robust to variations of the victim system location, the loudness of the legitimate user's vocalizations, etc.
To overcome the above issues, we first augment the optimization process of ultrasonic backdoors by introducing randomness into the user's speech content and time. As such, \bd can be applied with a wide variety of configurations in a content-agnostic and synchronization-free manner. 
Moreover, we adopt the natural evolution strategies~\cite{wierstra2014natural} to estimate the gradient via querying the black-box SRS models.
We further improve the robustness of our attack by taking into consideration the ultrasound attenuation varying with distances as well as the loudness relationship between the recovered backdoor and the legitimate user speech.

\blue{We validate \bd on seven representative SRS models, including ECAPA-TDNN~\cite{2020ecapatdnn}, Pyannote~\cite{bredin2020pyannote}, U-Level~\cite{xie2019utterance}, WavLM-Xvec~\cite{chen2022wavlm}, SpeakerNet~\cite{koluguri2020speakernet}, D-vector~\cite{wan2018generalized}, Sinc-Xvec~\cite{chung2020in} along with two typical speech datasets (VoxCeleb1~\cite{nagrani2017voxceleb} and LibriSpeech~\cite{panayotov2015librispeech}). Results show that \bd is effective in deceiving SRS and robust to various impact factors, such as different backdoor duration, speakers, speech content, and SRS models. Meanwhile, we also conduct a series of experiments to examine \bd's performance in the physical world and results demonstrate that \bd can work well under varying environments, attack distances, angles, and recording devices.
In addition, we examine \bd's resistance to seven representative audio signal preprocessing- and inaudible voice attack-based defenses, under the naive and adaptive adversary attack settings. Our defense experiments justify \bd can directly bypass almost all defenses, and successfully survive challenging median filter by utilizing adaptive attack strategy.
To mitigate this new threatening attack, we discuss several potential countermeasures and reveal its limitations, which may inspire future works.
}

In summary, our contributions are listed as follows:

\begin{itemize}[leftmargin=*,itemsep=0pt,partopsep=0pt,parsep=\parskip]
    \item We propose a new type of inaudible backdoor attack framework against speaker recognition systems via adversarial ultrasound and call for attention to a new attack surface.
	
    \item We introduce a collection of augmentation mechanisms to optimize the ultrasonic backdoor, which enables our attacks to maintain effective regardless of the victim speaker, speech content, speech volume or attack distances. 

    \item We propose efficient frequency sparsification, pre-compensation, and single-sideband modulation techniques to mitigate the signal distortion of ultrasonic attacks.
	
    \item \blue{We conduct extensive experiments to validate \bd's effectiveness under various configurations in both the digital and real-world scenarios, as well as its robustness against seven representative defenses.}

    \item \blue{We provide discussions on several potential countermeasures against this newly discovered threat, including using actively ultrasound carrier canceling mechanism and adopting sound field-based authentication.}
    
\end{itemize}


\section{Background and Threat Model}
\begin{figure}[t]
	\centering
	\includegraphics[width=0.45\textwidth]{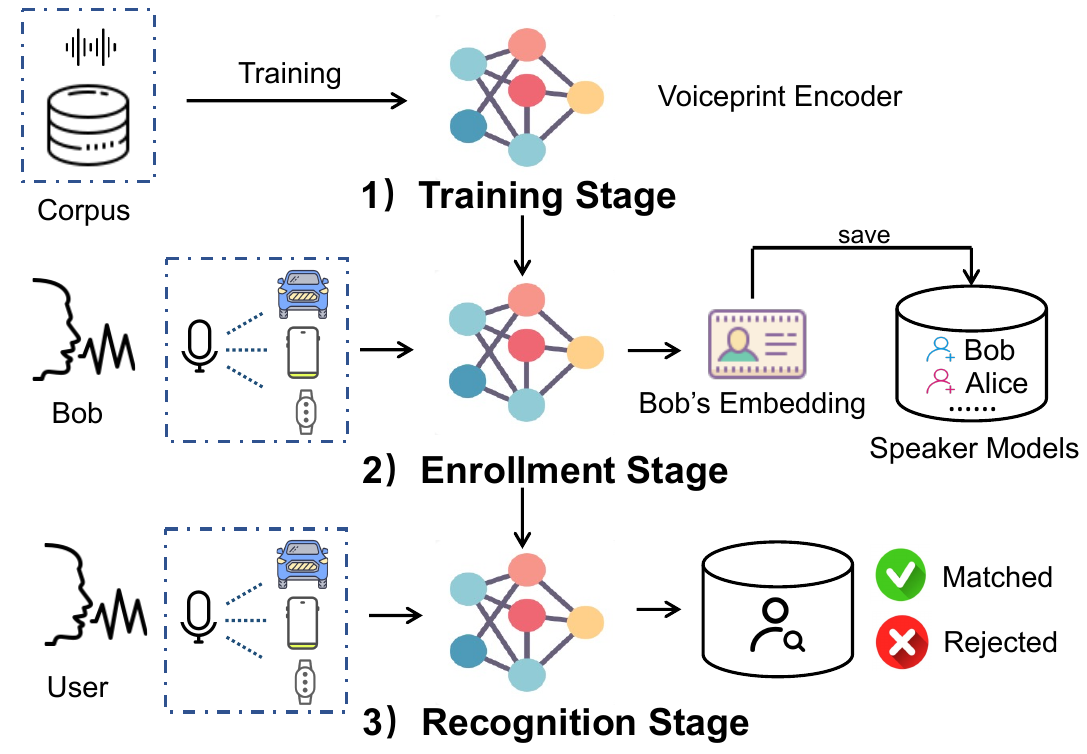}
	\caption{\label{fig:SRS}An illustration of a typical automatic speaker recognition system (SRS), which consists of three main stages, (1) training stage, (2) enrollment stage, and (3) recognition stage.}
\end{figure}

In this section, we first provide the background on automatic speaker recognition systems (SRS) and inaudible voice attacks. Then we present our threat model.

\subsection{Automatic Speaker Recognition System}
Speaker recognition systems model humans' voice characteristics (i.e., ``voiceprint''), to identify different speakers~\cite{jajodia2011encyclopedia}. 
A typical SRS consists of three stages: 1) training stage, 2) enrollment stage, and 3) verification stage. At the training stage, the SRS transforms the raw input audio sample into a feature vector, where candidate feature extraction algorithms include Mel-Frequency Cepstral Coefficients (MFCC)~\cite{sood2021speech}, Spectral Subband Centroid (SSC)~\cite{thian2004spectral} and Perceptual Linear Predictive (PLP)~\cite{hermansky1990perceptual}. Then an encoder is utilized to extract a low-dimensional representation (i.e., the unique voiceprint of a user) from the feature vector. Generally, the service providers of SRSs gain full control of the training process. At the enrollment stage, a user creates his voiceprint by the pre-trained model and registers as a legitimate user. Specifically, the user provides multiple audio samples, which are content-dependent or content-independent, to complete the process of enrollment. Such voiceprint is stored in the SRS and will be used to verify the user's identity during the verification stage. 
Through statistical modeling and initial exploitation of neural networks, researchers proposed the representative I-vectors~\cite{dehak2010front} based on the GMM-UBM and X-vectors~\cite{snyder2018x} based on the DNN, respectively. 
With the rapid development of deep learning techniques, SRSs have evolved into various state-of-the-art models, such as ECAPA-TDNN~\cite{2020ecapatdnn} and Pyannote~\cite{bredin2020pyannote}. In addition to model enhancements, PROLE Score~\cite{he2022ok} presents a content-related measurement and provides insights on voiceprint distinctiveness.

Speaker recognition can be classified into speaker identification/verification. Identification aims to determine from which of the registered speakers a given utterance comes, while speaker verification (SV) corresponds to accepting or rejecting the identity claimed by a speaker~\cite{Furui2008SRS}. 
Furthermore, the identification can be categorized into close-set identification (CSI) and open-set identification (OSI).

\textbf{Close-set identification (CSI).} A CSI system assumes that the speaker being verified belongs to a close set of enrolled users. To differentiate which (valid) user the speaker is, the speaker's voiceprint is compared with each enrolled speaker model for deriving the respective similarity score. The index of the highest score is considered as the speaker's identity.

\textbf{Open-set identification (OSI).} Different from the CSI, an OSI system takes into account that the speaker being verified may not belong to the set of enrolled users. Thus the system can be deployed in the wild with any possible intruders. The OSI also compares the speaker with each enrolled speaker model and obtains the identity with the highest similarity. Next, it further compares the highest score with a pre-defined threshold. If the score exceeds the threshold, this speaker is accepted as a legitimate user or vice versa.

\textbf{Speaker verification (SV).} An SV system also considers that the speaker being verified can be an intruders to the enrolled users. But it only holds a single enrolled speaker model, which can be regarded as a special OSI system. Therefore, SV compares the speaker's voiceprint with the claimed speaker model, and determines whether the speaker is legitimate or not~\cite{he2022ok}.

We will evaluate the attack performance of \bd in above three tasks in Sec.~\ref{sec:evaluation}.

\begin{figure}[t]
	\centering
	\includegraphics[width=0.45\textwidth]{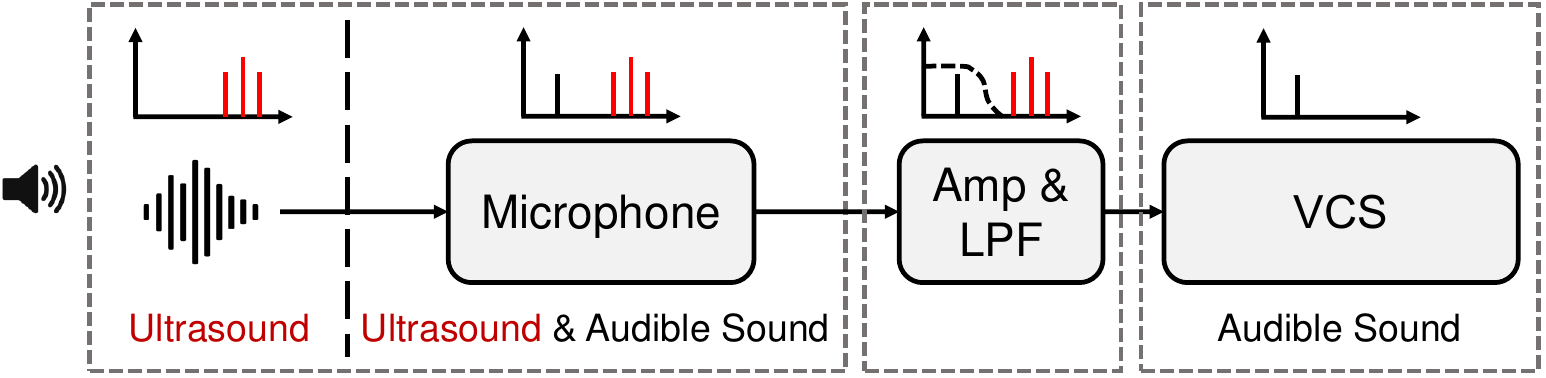}
	\caption{\label{fig:dolphinattack_pipeline}The general workflow of inaudible voice attacks. Attackers first modulate voice signals (i.e., baseband) on the ultrasound carrier (e.g., f$>$20~kHz). Due to the nonlinearity effect of microphones, such inaudible voice signals would be demodulated and recovered to low-frequency baseband signals, which can pass through the low-pass filter and further be interpreted by voice controllable systems (VCSs).}
\end{figure}

\subsection{Inaudible Voice Attacks}
Inaudible voice attacks inject voice commands to voice controllable systems by exploiting the nonlinearity effect of microphones while being entirely imperceptible~\cite{zhang2017dolphinattack,roy2018inaudible, ji2022capspeaker, yan2020surfingattack,sugawara2020light,wang2022ghosttalk}. Fig.~\ref{fig:dolphinattack_pipeline} presents the general workflow of inaudible voice attacks. First, malicious attackers modulate voice signals (also named baseband signals) on ultrasonic carriers (e.g., f$>$20~kHz). The typical modulation scheme is amplitude modulation (AM). Then, microphones will demodulate these ultrasound signals and output low-frequency baseband signals due to their nonlinearity effect. Finally, a low-pass filter will remove the ultrasound component of the recovered signals. Thus the demodulated voice signals can be almost the same as the normal ones, which makes the detection nontrivial, especially since it appears after the microphone module. Given that the modulated ultrasonic signals are carried above 20~kHz, inaudible voice attacks are highly imperceptible to human users.

\subsection{Threat Model}
\textbf{Knowledge.}
Given that the parameters or structures of most commercial SRSs are unavailable to the users, to make our attack practical, we focus on the black-box setting and demonstrate the feasibility of our enrollment-stage attack. Specifically, attackers have no knowledge of the target SRS, such as the gradient information or other metadata, which poses challenges for crafting backdoors. 
Moreover, we assume that the attacker can secretly record or collect audio samples of the victim in advance.

\textbf{Capability.}
First, malicious attackers can query the target SRS (e.g., by renting a smartphone or smart speaker of the same model) with generated backdoors using ultrasound modulation. Second, we assume that attackers can play ultrasonic backdoors physically close to the user during enrollment. In this case, both the ultrasonic backdoor and the victim's voice would be captured by the target SRS. Third, we envisage attackers deploying the ultrasonic transmitter covertly, where the victim SRS device is located in its attack range.


\section{Attack Investigation}\label{sec:investigation}
\subsection{Failure of Direct Inaudible Voice Attacks}
\begin{figure}
    \centering
    \includegraphics[width=0.45\textwidth]{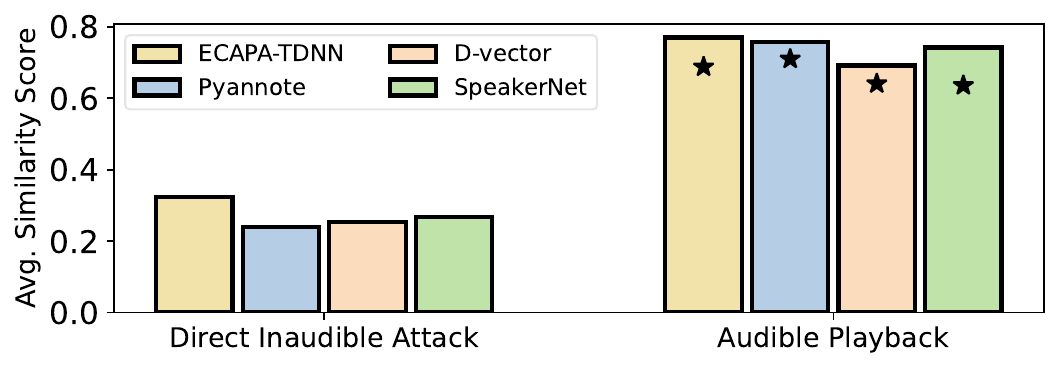}
    \caption{\label{fig:preliminary_asv}Average voiceprint similarity scores by the direct inaudible voice attack and the audible playback approaches. ``$\bm{\star}$'' means the threshold of the corresponding SRS model.}
\end{figure}

Motivated by the increasing prevalence of inaudible voice attacks~\cite{zhang2017dolphinattack,roy2018inaudible, ji2022capspeaker, yan2020surfingattack,sugawara2020light,wang2022ghosttalk} that can manipulate automatic speech recognition systems without being noticed, we propose that pre-recorded samples of a victim's voice could also be modulated and emitted using similar imperceptible techniques to successfully bypass SRSs.
We randomly selected six speakers from the Librispeech dataset and 30 sentences from each individual. All these utterances were launched by both direct inaudible voice attack and audible playback for explicit comparison. We employed a 25~kHz ultrasonic transducer array as the transmitter and two recording devices (i.e., Google Pixel and OPPO Reno5) 30~cm away to capture the baseband (i.e., backdoor) from the AM signal. Similarly, we used a JBL loudspeaker as the audible sound source. 
We evaluated all voice samples on four well-trained SRS models: ECAPA-TDNN, Pyannote, D-vector, and SpeakerNet. We calculated the similarity of each sample to its original voiceprint using cosine similarity scoring, and the average score of all samples was obtained. Our results, shown in Fig.~\ref{fig:preliminary_asv}, demonstrate that direct inaudible voice samples achieve significantly lower similarity scores (average: 0.240$\sim$0.323, all below the threshold ``\bm{$\star$}'') compared to audible playbacks (average: 0.692$\sim$0.770, all over the threshold ``\bm{$\star$}'') across the SRS models, indicating that directly emitting victim's voice through inaudible voice attacks are likely to be denied by SRSs.


We are driven to understand why direct inaudible voice attacks ill-perform in deceiving the speaker recognition tasks. Our further investigation reveals there are differences between such an attack and audible playback, which create typical challenges that impede applying direct inaudible voice attacks to manipulate voiceprint authentication:

\begin{itemize}[leftmargin=*,itemsep=0pt,partopsep=0pt,parsep=\parskip]
    \item \emph{Voiceprint Distortion:} Signal distortion leads to impaired voiceprint. 
    
    \item \emph{Processing Instability:} Complex processing and sophisticated hardware pipeline introduce instability.
\end{itemize}

\subsection{Voiceprint Distortion}\label{sec:signal_distortion}
\begin{figure}[!t]
    \vspace{-5pt}
    \centering
    \subfigure[Original Speech]{
    \begin{minipage}[t]{0.3\linewidth}
    \centering
    \hspace{-0.25\linewidth}
    \includegraphics[trim=0mm 0mm 0mm 0mm, width=1.05\textwidth]{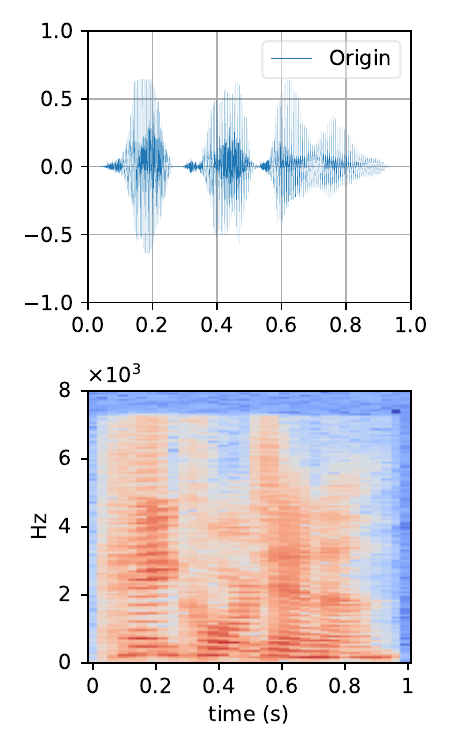}
    \end{minipage}%
    }%
    \subfigure[Playback Speech]{
    \begin{minipage}[t]{0.3\linewidth}
    \centering
    \hspace{-0.25\linewidth}
    \includegraphics[trim=0mm 0mm 0mm 0mm, width=1.05\textwidth]{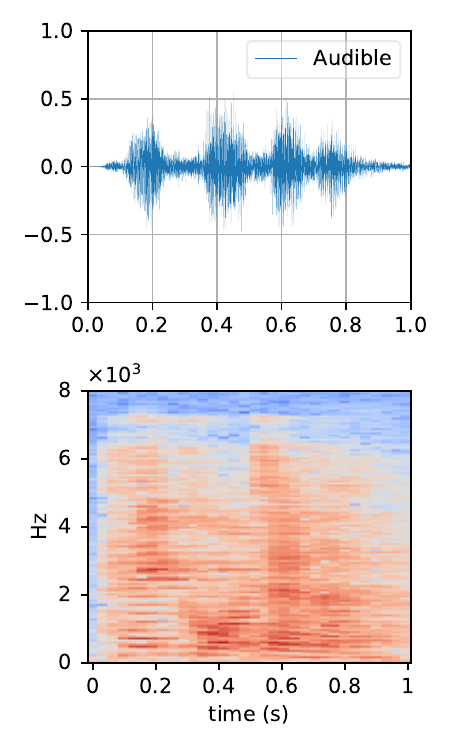}
    \end{minipage}%
    }%
    \subfigure[Inaudible Speech]{
    \begin{minipage}[t]{0.3\linewidth}
    \centering
    \hspace{-0.25\linewidth}
    \includegraphics[trim=0mm 0mm 0mm 0mm, width=1.05\textwidth]{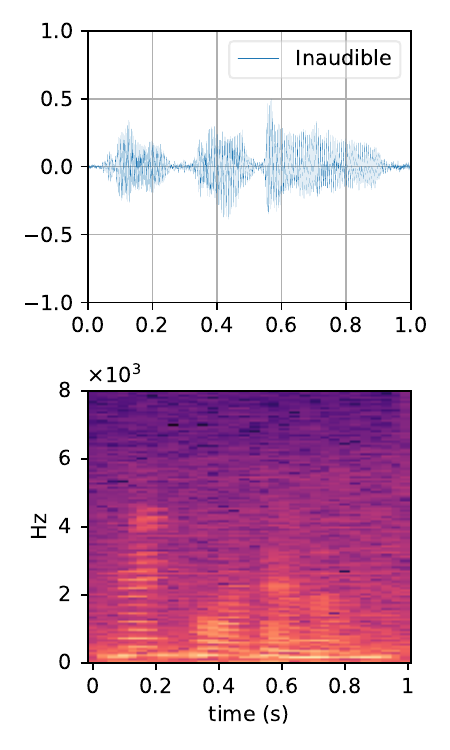}
    \end{minipage}
    }%
    \centering
    \caption{\label{fig:comparison_spec}Comparison of the (a) original speech and its (b) audible playback as well as (c) the inaudible voice attack.}
\end{figure}

Recent studies have shown that the nonlinear demodulation process of the inaudible voice attack follows a self-convolution operation~\cite{roy2018inaudible}. Such an endogenous mechanism of microphones' nonlinear loophole creates signal distortion, which includes a concentration of the recovered baseband's energy aggregated in the low-frequency band (e.g., sub-50~Hz), as well as intermodulation between various frequencies. Collectively, we refer to these phenomena as signal distortion.
To visually compare the differences between audible playbacks and direct inaudible voice attacks, we take one speech sample in Fig.~\ref{fig:comparison_spec}(a), and present its audible playback and direct inaudible voice attack versions in Fig.~\ref{fig:comparison_spec}(b),(c), respectively. Mainstream SRSs use spectral features as the input of voiceprint models, and we observe that the audible playback corresponds perfectly to the original audio. However, the spectrum features of the recovered audio by the direct inaudible voice attack are distinct from the original audio; namely, the baseband's frequency pattern (i.e., pitch contours) is considerably modified. Previous works have demonstrated that pitch changes can result in noticeable performance degradation on speaker verification/recognition tasks~\cite{ai2019improvement,o2022evaluating,tull1996analysis,wagner2017infected}. Moreover, pitch-shift operations can even be applied to disguise voices against SRSs~\cite{zheng2020automatic,tavi2022improving}. Pitch changes barely affect the speech recognition models to infer the phoneme sequence, making direct ultrasound modulation of voice commands still recognizable. In contrast, SRS models cannot extract the correct voiceprint when the input spectral features are distorted.


\begin{figure}
    \centering
    \includegraphics[width=0.4\textwidth]{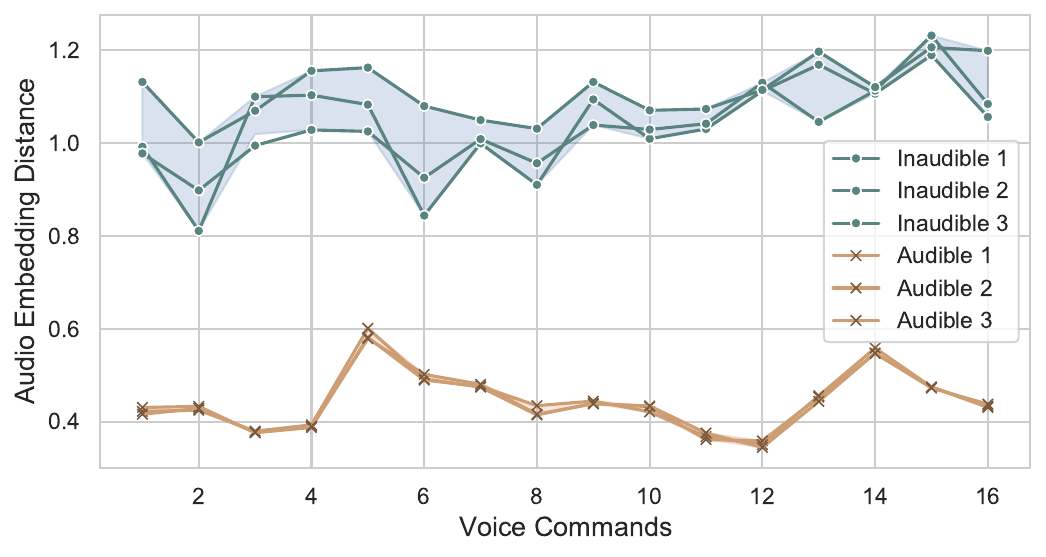}
    \caption{\label{fig:audio_quality_stability}Audio similarity of the corresponding original audio by repeating audible playbacks and inaudible voice attacks three times.}
    \vspace{-10pt}
\end{figure}

\subsection{Attack Instability}
Existing audible-band speech spoofing attacks against SRSs, e.g., replay attacks, are launched by typical loudspeakers, whereas emitting the ultrasonic backdoor relies on the signal generator for real-time amplitude modulation and the power amplifier for a wider attack range. Thus, more hardware imperfection is introduced. To quantify the impact of such a complex pipeline compared to audible playback, we controlled variables such as the relative distance of attack and fixed recording devices. We respectively performed audible playback and inaudible voice attacks by launching the same commands and recording as multiple audio samples, with each operation repeated three times. Applying CDPAM~\cite{manocha2020differentiable}, a deep learning-based audio similarity metric tool, we derived the audio embedding distance between each recorded audio and its origin. Note that a higher distance (e.g., $>$0.8) means that more severely the audio is altered from the origin due to signal distortion. Fig.~\ref{fig:audio_quality_stability} shows that the distances of audible playbacks to their origin are significantly closer than the inaudible, in line with the results given in  Fig.~\ref{fig:preliminary_asv} and Fig.~\ref{fig:comparison_spec}.
Furthermore, we observe that the distances of same commands by audible playback are almost identical, while there were remarkable deviations (the shaded area) among the inaudible ones. Namely, a series of processing and hardware imperfection leads the inaudible voice attacks to relatively low stability.

\begin{figure*}[!t]
	\centering
	\includegraphics[width=0.99\textwidth]{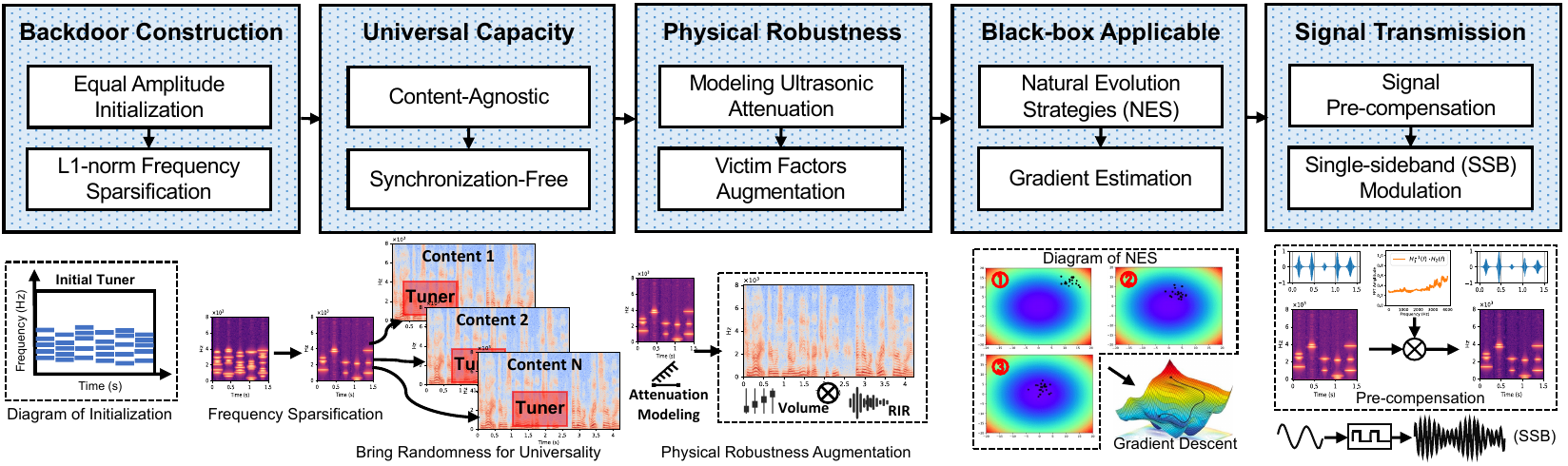}
	\caption{\label{fig:design-overview}\small The workflow of \bd. An initial adversarial ultrasound would be optimized to meet the optimization requirements of power-robust, content-agnostic, and synchronization-free simultaneously.}
	\vspace{-10pt}
\end{figure*}

\section{System Design}

Existing adversarial example attacks generally mislead the recognition results of the SRS by adding slight perturbations to the original audio, which are still perceptible to human beings. However, our backdoor attack emits triggers modulated on adversarial ultrasound at the enrollment stage, which are beyond the audibility range of humans and exhibit the advantage of high imperceptibility. Our goal is to poison the user's voiceprint during enrollment, allowing both legitimate users and malicious attackers to pass the target SRS with relatively high confidence scores at the recognition stage. 
To achieve this, we carefully design \bd's workflow in Fig.~\ref{fig:design-overview} that accomplishes the following goals. 

\begin{itemize}[leftmargin=*,itemsep=0pt,partopsep=0pt,parsep=\parskip]
	\item \textbf{Backdoor Construction (G1).} The backdoor shall be constructed efficiently, and meanwhile, its frequency components are simplified during optimization to mitigate signal distortion when performing physical attacks.
	\item \textbf{Universal Capacity (G2).} Malicious attackers have no prior knowledge of the exact content of the user's audio samples or when the user vocalizes during enrollment. Thus, the ultrasonic backdoor shall be applied to arbitrary content or starting time of the user speaking out.
	\item \textbf{Physical Robustness (G3).} The ultrasonic backdoor is expected to perform successful attacks regardless of the ultrasound attenuation at different attack distances and user loudness levels or environments.
	\item \textbf{Black-box Applicable (G4).} In most cases, attackers have no access to the gradient information of the commercial SRS models while attacking. Enabling black-box attack capability is necessary to make the attack practical.
	\item \textbf{Signal Transmission (G5).} Attackers should further mitigate the signal distortion challenges that exists during signal transmission by effective signal compensation and modulation methods.
\end{itemize}


\subsection{Problem Formalization}
The overall goal of our attack is to maximize the probability of a backdoor being recognized as a legitimate user by the target SRS while ensuring that the victim can still pass the recognition. To this end, we formulate the generation of backdoors as an optimization problem as follows.

Given an original audio sample $x$ of the victim, we aim to craft a robust baseband signal with only a few frequencies considering realistic constraints mentioned in Sec.~\ref{sec:investigation}, formalized as ${p} = \sum\nolimits_{m = 1}^M{\sum\nolimits_{n = 1}^N {{\mathcal{A}_n} \cdot \sin (2\pi {\mathcal{K}_n}{T_m})} }$. By finding the amplitude matrix $\mathcal{A}$ and frequency matrix $\mathcal{K}$, the target SRS can accept the voiceprint of both the victim and \bd.
\begin{equation}\label{eq1}
    \small
    \mathcal{L} (x,p) = - {\alpha_{1}{S(\widetilde{x},~}}{\rm{X}}_\textit{victim})-\alpha_{2}{S({\widetilde{x}},~}{{\rm{X}}_{\textit{\bd}}})
\end{equation}
where $\mathcal{L}(x,p)$ represents the loss function for deriving the target backdoor trigger $p$; ${\widetilde{x}}={x}+{p}$ denotes the victim's voice $x$ and the ultrasonic backdoor $p$ are captured by the recording device during enrollment. $X_\textit{victim}$ and $X_{\textit{\bd}}$ are the unpolluted voiceprints of the victim and the backdoor, respectively; $\alpha_{1,2}$ are the weights, where a larger value corresponds to the higher optimization importance. $S(\cdot)$ means the cosine similarity scoring module to calculate the similarity between two voiceprints. $M$ is the number of segments we slice the $p$ into, while $N$ denotes the number of frequency points contained in every segment. We can gain the ultimate backdoor trigger $\hat{p}$ by solving 
$$\hat{p}=\underset{\mathcal{A,K}}{\operatorname{argmin}}\mathcal{L} (x,p)$$

\subsection{Backdoor Construction}\label{example-construction}
\textbf{Equal Amplitude Initialization.}
The relative energy of our ultrasonic backdoor and victim's voice samples significantly affects the convergence speed during generating \bd. Our empirical experiment shows that optimizing a trigger that meets the conditions of Eq.~\ref{eq1} is easier when \bd and victim utterances have similar volume in the time domain. Therefore, we introduce equal amplitude initialization to facilitate the process of backdoor optimization, where items of the amplitude matrix $A$ are identical and averaged by the maximum amplitudes of the secretly collected victim voice samples (denoted as $G$), shown as ``Initial Tuner'' in Fig.~\ref{fig:design-overview} bottom left.
\begin{equation}\label{eq:amp_initial}
    \small
    {\mathcal{A}} \gets avg[\sum_{i=1}^{G} {{max(x_\textit{victim})}}]
\end{equation}

\textbf{L1-norm Frequency Sparsification.}
Unlike initializing the items of the amplitude matrix $\mathcal{A}$ identically using Eq.\ref{eq:amp_initial}, we generate the frequency matrix $\mathcal{K}$ initially with multiple frequency components as they can poison the victim's voiceprint more effectively. However, considering the perspective of real-world attacks, we need to simplify the trigger. To achieve this, we adopt a dense-to-sparse strategy using L1-norm penalty\cite{tibshirani1996regression}, instead of crafting substantial \bd candidates with sparse frequencies and selecting the best one. 
L1-norm regularization is commonly used to prevent model over-fitting, which automatically sets unimportant parameters to zero during the training process. Drawing inspiration from the demonstrated model parameter simplification using L1-norm, we apply this idea to our backdoor by retaining the most significant frequency components that poison the victim's voiceprint. We materialize frequency sparsity based on Eq.~\ref{eq1} as follows:
\begin{equation}\label{eq:freq_sparse}
    \small
    \mathcal{L} (x,p) = - {\alpha_{1}{S(\widetilde{x},~}}{\rm{X}}_\textit{victim})-\alpha_{2}{S({\widetilde{x}},~}{{\rm{X}}_{\textit{\bd}}}) + \alpha_{3}L_{1}(\mathcal{A})
\end{equation}
where $\alpha_{3}$ means a hyper-parameter that weights the $L_{1}(\mathcal{A})$ penalty term. During the optimization, multiple items of $\mathcal{A}_{n}\cdot sin(2\pi\mathcal{K}_{n}\mathcal{T}_{m})$ will not work due to $\mathcal{A}_{n}$ automatically reducing to zero, as the illustration ``Frequency Sparsification'' shown in Fig.~\ref{fig:design-overview}. Notably, we do not optimize the frequency matrix $\mathcal{K}$ simultaneously because it can lead to convergence oscillation. 

\subsection{Universal Capacity}\label{universal-capacity}
\textbf{Content-Agnostic.}\label{Content-Agnostic}
Mainstream speaker recognition systems are typically text-independent, i.e., the voiceprints extracted from different utterances of a given user may not vary significantly. However, it is not trivial to ensure that the poisoned voiceprints remain consistent when a trigger is applied to different victim voice samples. As victims' utterances can vary across sessions during enrollment, it is not practical to regenerate the trigger each time the victim's speech content changes. To address this issue, we introduce a parameter $v$ in the backdoor generation process. By solving Eq.~\ref{eq4}, we optimize the same backdoor trigger for different utterances of the victim in $V$. Notably, we have validated in advance that \bd can be effective on each utterance of $V$, i.e., achieving a content-agnostic backdoor attack.
\begin{equation}\label{eq4} 
    \small
	\mathcal{L}(x,p) = -\alpha_{1}S({\widetilde x_{v}}{\rm{,}}{{\rm{X}}_\textit{victim}}) - \alpha_{2}{S}{(\widetilde x_{v}}{\rm{,}}{{\rm{X}}_{\textit{\bd}}})  + \alpha_{3}L_{1}(\mathcal{A})
\end{equation}
where ${\widetilde x_{v}} = {x_v} + p, v\in{V}$.

\textbf{Synchronization-Free.}\label{Synchronization-Free}
Most existing adversarial example (AE) attacks are feasible only in specific scenarios, where they assume that the attacker can predict the audio vocalized by the victim and play the AEs synchronously at fixed time points to carry out the attack. It is hard to implement since the audio signal is time-sensitive. Motivated to craft backdoors more robust in physical attacks, we randomly select the starting point, from which the backdoor superimposed on the victim's original audio signal, composing a preset time range $T$ to simulate an attacker launching attacks at any time at the enrollment stage.
Within the range $T$, we introduce a parameter $t$ that dynamically changes during the backdoor generation. Integrating the above optimization objective ``Content-Agnostic'', this process is visually rendered with several victim's content clips and a given \bd shifting on them in Fig.~\ref{fig:design-overview}. We obtain the final optimization function by solving the following Eq.~\ref{eq5}.
\begin{equation}\label{eq5}
    \small
	\mathcal{L}(x,p) = -{\rm{\alpha_{1}S(}}{\widetilde x_{v,t}}{\rm{,}}{{\rm{X}}_\textit{victim}}) - \alpha_{2}{\rm{S(}}{\widetilde x_{v,t}}{\rm{,}}{{\rm{X}}_{\textit{\bd}}})  + \alpha_{3}L_{1}(\mathcal{A})
\end{equation}
where ${\widetilde x_{v,t}} = {x_v} + {\rm{shift}}(p,t), v\in{V}, t\in{T}$. The shift operation mimics \bd $p$ can be randomly superimposed on arbitrary victim speech $x_v$ within the preset $T$.

\subsection{Physical Robustness}\label{Physical-Robustness}


\textbf{Modeling Ultrasonic Attenuation.}
The high-frequency property of ultrasound leads it to be more easily attenuated. Besides, the energy consumption is also related to the air viscosity, temperature and humidity~\cite{Sound}. Therefore, the energy received by victim devices varies with the attack launching locations. The optimization phase shall consider the relative energy variation between the backdoor demodulated from ultrasound (i.e., affected by changing attack distances) and the legitimate user's speech. Such attenuation is a power law frequency-dependent acoustic attenuation~\cite{zhang2021eararray}, and can be expressed as:
\begin{equation}\label{equ:dis_filter}
    \small
    p^d = p\cdot e^{-a_0{\omega_c}^n d},~n\in[1,2]
\end{equation}
where $a_0$ is a medium-dependent attenuation parameter, $\omega_c$ is the carrier's frequency configured as 25~kHz, and $d$ is random within [0.3m, 2m] in our case.

\textbf{Victim Factors Augmentation.}
Moreover, the victims' speech volume and their environment (i.e., the presence of reverberation) are also various and unpredictable. These factors make it challenging for a digitally well-crafted \bd to maintain effectiveness in real-world scenarios.
To tackle issues of victim factors, we propose two techniques: relative volume augmentation and the room impulse response (RIR) simulation, which combined with ``ultrasound attenuation'' are expressed as ``physical robustness augmentation'' in Fig.~\ref{fig:design-overview}.
The former involves a parameter $beta$ that introduces randomness into the loudness relationship between the adversarial ultrasound and the victim's voice during the backdoor generation process. We set the range of $\beta$ by restricting the relative power ${\text{ratio}} = E(\beta\cdot{x}_{\textit{victim}})/E({p})$ to a reasonable level, e.g., $\beta \in [0.5, 2]$. $E( \cdot )$ denotes the power of audio.
The latter aims to effectively poison the voiceprint regardless of the environment influence. We utilize the random RIR clips in the Aachen impulse response (AIR) database~\cite{jeub2009binaural}, which includes small, medium, and large rooms, for human speech enhancement.
\begin{equation}\label{eq7}
    \small
    \mathcal{L}(x,p) = -\alpha_{1}{S(}\widetilde x^{d}_{\beta,v} ,{{\rm{X}}_\textit{victim}}) - \alpha_{2}{S(}\widetilde x^{d}_{\beta,v} ,{{\rm{X}}_{\textit{\bd}}})  + \alpha_{3}L_{1}(\mathcal{A})
\end{equation}
where ${\widetilde x^{d}_{\beta,v}} = \beta\cdot x_v + {p^d}$.


\subsection{Black-box Applicable}\label{Black-box}
Gradient information in Eq.~\ref{eq7} is easy to obtain in the white-box settings. By contrast, in the black-box settings, the adversary cannot update $\delta$ in the same manner due to lack of the gradient information.
To launch a practical black-box attack, we adopt natural evolution strategies (NES)~\cite{wierstra2014natural,deng2022fencesitter}, which renders an efficient strategy to estimate the gradient within limited query times and depict the diagram of it evolution in Fig.~\ref{fig:design-overview}. NES assesses $\nabla_{\mathcal{(A,K)}}\mathcal{L}$ in Eq.~\ref{eq7} based on the similarity scores $S(\tilde{x}, X)$ that can be obtained by querying the black-box SRS.
Specifically, given an initiated ultrasonic backdoor pulse $x$, we search the better pulse by leveraging the Algorithm~\ref{algo:nes}
, where $\delta\sim N(0,I)$, $P$ is the dimension of a backdoor trigger $p$. Through symmetrically sampling to generate $\pm\delta_i$, where $i\in\{1,...,n\}$, we employ limited query to gain $\frac{1}{2n\sigma}g$, which is the estimation gradient information of $\nabla\mathcal{L}(x,p)$. Empirically, we set the number of samples $n$ = 15 and $\sigma$ = 0.08. The whole algorithm of \bd is summarized in Algorithm 1.

\begin{algorithm}[t]
	\caption{NES Gradient Estimation}
	\label{algo:nes}
	\LinesNumbered
	\small
	\KwIn{Query the target SRS by the backdoor $p$ and victim sample $x$.}
	\KwOut{Estimation of $\nabla{\mathcal{L}(x,p)}$}
	\KwData{search variance $\sigma$, number of samples n, backdoor dimensionality P=M$\times$N}
		${g} \gets 0_{P}$\\
		\For{$1$ to $n$}
		{
			$\delta_{i} \gets N(0_{P}, I_{P\cdot P})$\\
			$g \gets g + \mathcal{L}(x,p+\sigma\cdot\delta_i)\cdot \delta_i$\\
			$g \gets g - \mathcal{L}(x,p-\sigma\cdot\delta_i)\cdot \delta_i$\\
		}
		\KwRet{$\frac{1}{2n\sigma}g$}
	\normalsize
\end{algorithm}

\begin{figure}[!t]
	\centering  
	\subfigure[]{   
	\begin{minipage}[t]{0.18\textwidth}
		\centering
		\includegraphics[width=1\textwidth]{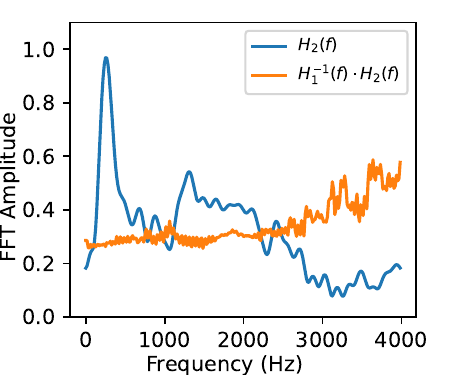}
	\end{minipage}
	}\hspace{-0.03\textwidth}
	\subfigure[]{   
	\begin{minipage}[t]{0.3\textwidth}
		\centering
		\includegraphics[width=1\textwidth]{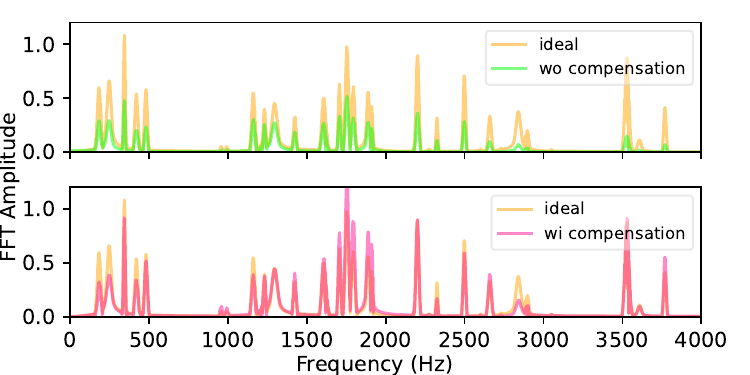}
	\end{minipage}
	}
	\caption{Best viewed in color. (a) Diagram of the microphone's frequency response $H_r(f)$ and the inverse filter $H_{1}^{-1}(f)\cdot H_2(f)$. (b) Fourier transforms of the ideal \bd: yellow, without compensation: green, and with compensation: pink.}    
	\vspace{-5pt}
	\label{fig:IR_recon}    
\end{figure}

\subsection{Signal Transmission}\label{Signal Transmission}
\textbf{Signal Pre-compensation.}
Microphones equipped by smart devices present irregular frequency responses to inaudible voice attacks~\cite{li2023learning}, i.e., the backdoor goes through a signal-distortion nonlinear demodulation that has been demonstrated in Sec.~\ref{sec:signal_distortion} and Fig.~\ref{fig:comparison_spec}.
To overcome the challenge, we pre-compensate the backdoor signal before emission. We first analyze the frequency response of the recorder's microphone ${H_r}(f)$ and the frequency response between the transmitter and the recorder ${H_t}(f)$. Then, an attack signal is compensated to counter the distortion with an inverse filter $F$ before modulation. $F=h_r^{ - 1}(t) * {h_t}(t)$ where ${h}(t)$ converts to ${H}(f)$ by Fourier transform. ${H_r}(f)$ and ${H_t}(f)$ are analyzed on several recorders and respectively averaged for compensation~\cite{infomasker2023}.

\begin{algorithm}[t]
	\caption{Backdoor Trigger Generation}
	\label{algo1}
	\LinesNumbered
	\small
	\KwIn{The target SRS with a scoring module: $S$, the maximum epoch: $ maxEpoch$, the desired score of the fitness function: $J$, the learning rate: $\eta$, the preset time range: $T$, the set of victim's utterance samples $x_{vic}$: $V$} 
	\KwOut{The desired baseband of \bd}
	\textbf{Init} $\mathcal{A} \gets avg[\sum max(x_{vic})] $\\
	\textbf{Init} $\mathcal{K} \gets {rand\_freq}~(0,4~kHz) $\\
	\textbf{Init} ${p} \gets \text{\bd}(\mathcal{A,K})=\sum\limits_{m = 1}^M {\sum\limits_{n = 1}^N {\mathcal{A}_{n} \cdot \sin (2\pi \mathcal{K}_{n}\mathcal{T}_{m})} }$\\
	\For{$1$ to ${ maxEpoch}$}
	{
		${J} \gets 0$\\
            \For{$(v,t)$ in rand\_pick $\{V,T\}$}
            {
                $\widetilde x \gets x_{v} {\text{+}} \textit{shift}(p^d, t)$\\
                \For{$\beta$ in rand $(0.5,2)$}
                {
                  $\widetilde x \gets \beta\cdot x_{v} {\text{+}} \textit{shift}(p^d, t)$\\
                  $J \gets J {\text{-}}\alpha_{1} S({\widetilde x},{X_\textit{vic}}) {\text{-}} \alpha_{2} S({\widetilde x}, X_\text{\bd}) {\text{+}} \alpha_{3} L_{1}(\mathcal{A})$
                }
            Compute ${{\nabla}_\mathcal{A}}J$\\
            $\mathcal{A} \gets clip(\mathcal{A} {\text{+}} \eta  \cdot {\nabla_\mathcal{A}}J,[0,~1])$\\
            \If{$J \ge objScore$}{break}
            }
        }
	return $\text{\bd}(\mathcal{A,K})$
	\normalsize
\end{algorithm}

\textbf{Single-sideband Modulation.}
Distinct from traditional inaudible voice attacks~\cite{zhang2017dolphinattack,roy2018inaudible} that adopt double-sideband amplitude modulation (DSB-AM) to make the baseband (i.e., voice command) to an ultrasonic frequency beyond human auditory, which contains both upper and lower sidebands and the ultrasound carrier. 
In contrast, our imperceptible backdoor attack is designed to modulate the trigger using single-sideband amplitude modulation (SSB-AM). This technique removes one of the sidebands based on the Hilbert transform, resulting in a narrower bandwidth compared to DSB-AM. Thus, SSB-AM is more power-efficient than DSB-AM as it eliminates the redundant information in one of the sidebands, leading to a reduced transmission bandwidth and improved spectral efficiency. Furthermore, SSB-AM mitigates the intermodulation in signal distortion since the sideband frequency components are half-reduced.

Overall, the algorithm of \bd is described in Algorithm~\ref{algo1}, where we demonstrate the black-box setting-based optimization process of crafting \bd from scratch. To speed up the convergence, we employ Adam~\cite{kingma2014adam} to optimize the parameter $\mathcal{A}$ adaptively.


\section{Evaluation}\label{sec:evaluation}
In this section, we first describe our experiment setup. Then we examine the performance of \bd in terms of digital attacks and physical attacks under SV, CSI, and OSI scenarios along with multiple factors, including different attack duration, target speakers, etc. 

\subsection{Experiment Setup}\label{ssec:setup}

\textbf{Hardware} We implement our \bd prototype on a server with Ubuntu 18.04 Intel(R) Xeon(R) Gold 6240 CPU running at 2.60GHz and one GeForce RTX 3090 GPU. \blue{We employ the Keysight EXG signal generator~\cite{Keysight} and an ultrasound transducer array to modulate \bd as the baseband, as well as a power amplifier (NF HSA4015)~\cite{amplifierHSA4015} to enable long-range ultrasonic attack transmission. JBL loudspeakers are utilized to playback the victim's utterances and noises from the datasets below.}

\blue{\textbf{Dataset.} We adopt six widely-used human speech datasets to examine the performance of \bd. Particularly, we set the English VoxCeleb~\cite{nagrani2017voxceleb} and LibriSpeech~\cite{panayotov2015librispeech} as two main datasets as our default experiment configuration. We also consider the impact of different languages on \bd and evaluate it using AISHELL-1 (Chinese)~\cite{aishell_2017}, MLS Italian, MLS German, MLS French~\cite{Pratap2020MLSAL}. We randomly select 3 female and 3 male speakers as the candidate victims from each dataset, given in Tab.~\ref{tab:people}.} For each victim, we preset the attacker's ability to secretly record 3 victim's voice samples and launch 5 ultrasonic backdoors. Then, another unseen 20 utterance samples of each victim are used to test these backdoors' performance. 

\textbf{Target Model.} We validate \bd on seven representative end-to-end SRS models, including ECAPA-TDNN~\cite{2020ecapatdnn}, Pyannote~\cite{bredin2020pyannote}, U-Level~\cite{xie2019utterance}, WavLM-Xvec~\cite{chen2022wavlm}, 
SpeakerNet~\cite{koluguri2020speakernet}, D-vector~\cite{wan2018generalized}, ResNet34~\cite{chung2020in} along with two typical speech datasets (i.e., VoxCeleb1~\cite{nagrani2017voxceleb} and LibriSpeech~\cite{panayotov2015librispeech}). Particularly, our experiments are conducted under the black-box setting, i.e., we estimate the gradient instead of using explicit information. We choose ECAPA-TDNN as our main target model due to its relatively high performance among state-of-art TDNN-based systems. Notably, we set the threshold $\theta$ to 0.688 for ECAPA-TDNN according to the equal error rate (EER) on LibriSpeech (more details are listed in Tab.~\ref{tab:model_threshold}).

\begin{table}[t]
\scriptsize
\centering
\renewcommand\arraystretch{0.8} 
\setlength\tabcolsep{3.8pt}
\setlength{\abovecaptionskip}{0pt}%
\setlength{\belowcaptionskip}{0pt}%
\caption{detail of speakers used for evaluation}
 
\begin{tabular}{c|c|c|c|c|c|c}
\toprule
\textbf{People}& \multicolumn{1}{c}{\textbf{F1}} & \multicolumn{1}{c}{\textbf{F2}} & \multicolumn{1}{c}{\textbf{F3}} & \multicolumn{1}{c}{\textbf{M1}} & \multicolumn{1}{c}{\textbf{M2}} & \multicolumn{1}{c}{\textbf{M3}} \\ \midrule
\textbf{Librispeech} & 1580                                & 6829                                & 3570                                & 2830        & 7021        & 5105        \\ \midrule
\textbf{Voxceleb1}   & id10038                             & id10070                             & id10092                             & id10143     & id10176     & id10203     
\\ \midrule
\textbf{AISHELL-1}   & S0750	 &  S0761 &  S0770 & S0901 & S0912 & S0916
\\ \midrule
\textbf{MLS Italian}  & 280 & 1131 & 4009 & 428 & 646 & 6698
\\ \midrule
\textbf{MLS German}  & 278 & 1262 & 3588 & 91 & 144 & 1874
\\ \midrule
\textbf{MLS French}  & 2085 & 2154 & 2465 & 296 & 1406 & 2114
\\ \bottomrule                                        
\end{tabular}
\label{tab:people}
\end{table}

\color{black}
\begin{table}[t]
\centering
\renewcommand\arraystretch{0.8} 
\setlength\tabcolsep{3.8pt}
\setlength{\abovecaptionskip}{0pt}%
\setlength{\belowcaptionskip}{0pt}%
\caption{thresholds and EERs of different models}
\begin{tabular}{c|c|c|c|c}

\toprule
 & \multicolumn{2}{c|}{\textbf{Voxceleb1}} & \multicolumn{2}{c}{\textbf{Librispeech}} \\
{\textbf{Model}}         & \multicolumn{1}{c}{\textbf{Threshold}} & \multicolumn{1}{c|}{\textbf{EER(\%)}} & \multicolumn{1}{c}{\textbf{Threshold}} & \multicolumn{1}{c}{\textbf{EER(\%)}} \\ \midrule
{\textbf{ECAPA-TDNN}}   & 0.720     & 0.90    & 0.688       & 1.59    \\ \midrule
{\textbf{Pyannote}}      & 0.768     & 2.68    & 0.724       & 3.53    \\ \midrule
{\textbf{U-Level}}       & 0.777     & 3.06    & 0.789       & 2.41    \\ \midrule
{\textbf{WavLM-Xvec}} & 0.645     & 1.05    & 0.665       & 4.75    \\ \midrule
{\textbf{SpeakerNet}}    & 0.645     & 2.30    & 0.636       & 3.56    \\ \midrule
{\textbf{D-vector}}      & 0.777     & 11.91   & 0.641       & 7.05    \\ \midrule
{\textbf{ResNet34}}      & 0.675     & 2.11    & 0.656       & 5.42    \\ \bottomrule     
\end{tabular}
\label{tab:model_threshold}
\end{table}

\textbf{Evaluation Metrics.} We adopt the following metrics throughout the evaluation. (1) \emph{Attack Success Rate (ASR)} characterizes the rate at which the adversary successfully passes the recognition of the target SRS, i.e., the number of accepted samples over the total number of the adversary's test samples. (2) \emph{Accuracy (ACC)} characterizes the rate at which the target SRS correctly recognizes the legitimate user.

\subsection{Digital Attack Performance}
Under the digital attack scenario, the target SRS stores the average poisoned voiceprints, which are obtained by each victim's 3 speech samples embedded with the user-specific ultrasonic backdoor during enrollment. \blue{At the recognition stage, we feed the backdoor triggers and unseen victims' voice samples into the target SRS to examine their ASR/ACC performance, respectively.}

\begin{figure}[t]
	\centering
	\includegraphics[width=0.42\textwidth]{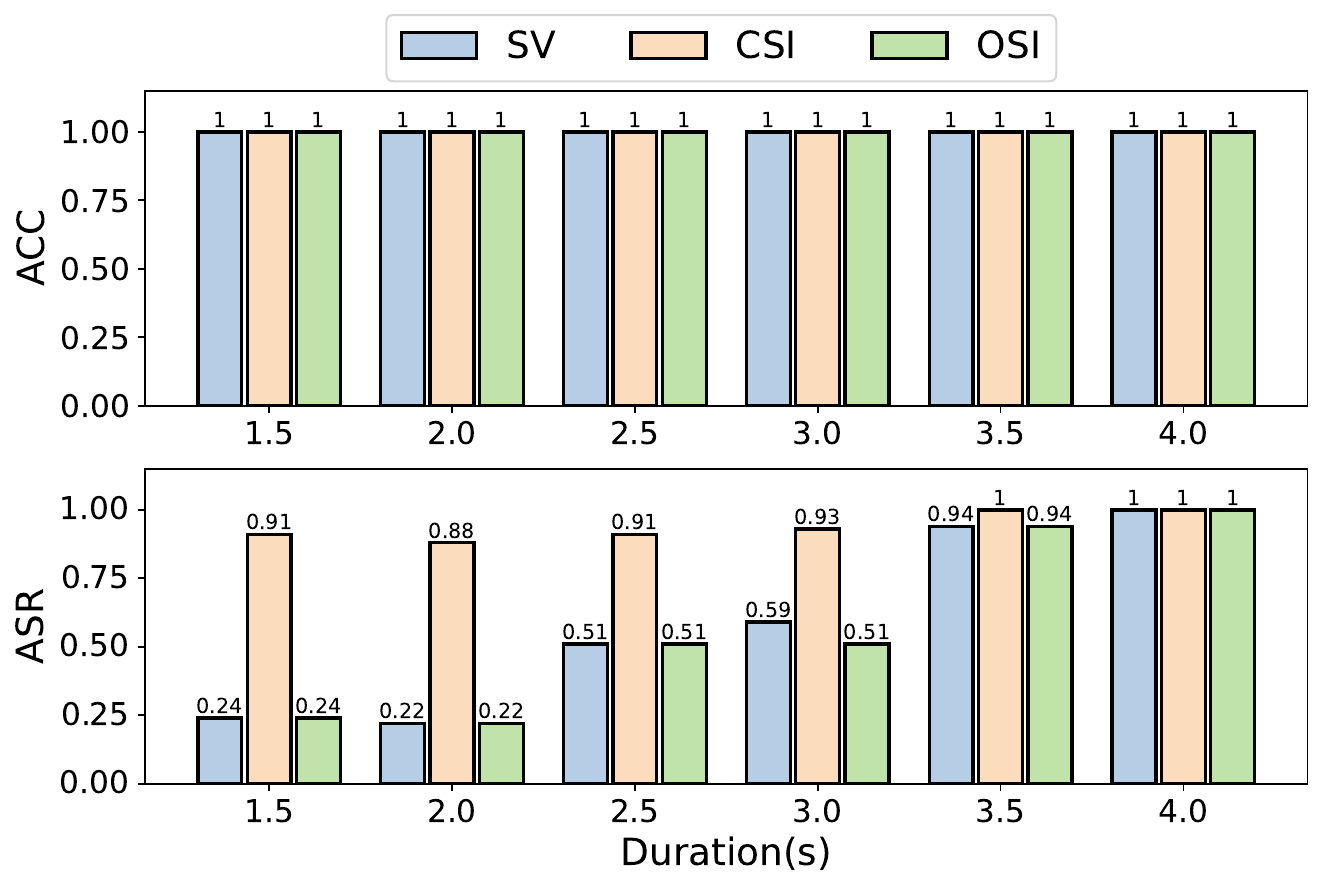}
	\caption{ASR/ACC performance of 6 \bd's duration (poisoning rate). The longer duration corresponds to the better ASR performance.}
	\label{fig:duration} 
\end{figure}

\begin{table*}[t]
	\scriptsize
	\centering
	\renewcommand\arraystretch{1} 
	\setlength\tabcolsep{2.5pt}
        \setlength{\abovecaptionskip}{0pt}%
        \setlength{\belowcaptionskip}{0pt}%
	\caption{Impact of different speaker recognition models.}
	\begin{tabular}{ccc|cccccccccccc|cccccccccccc}
		\toprule
		&  &  & \multicolumn{12}{c|}{\textbf{ECAPA-TDNN (\%)}}                                                 & \multicolumn{12}{c}{\textbf{PYANNOTE (\%)}}                                                 \\ \cline{4-27} 
		&  &  & \multicolumn{6}{c|}{\textbf{LibriSpeech}}          & \multicolumn{6}{c|}{\textbf{Voxceleb1}} & \multicolumn{6}{c|}{\textbf{LibriSpeech}}          & \multicolumn{6}{c}{\textbf{Voxceleb1}} \\
		&  &  & F1 & F2 & F3 & M1 & M2 & \multicolumn{1}{c|}{M3} & F1   & F2   & F3   & M1   & M2   & M3   & F1 & F2 & F3 & M1 & M2 & \multicolumn{1}{c|}{M3} & F1   & F2   & F3   & M1   & M2   & M3  \\ \midrule
		\multirow{2}{*}{\textbf{SV}}  & &ACC &100    &100    &100    &100    &94    & \multicolumn{1}{c|}{99}   &99      &90      &100      &93      &100      &100      &100    &100    &100    &100    &100    & \multicolumn{1}{c|}{95}   &100      &100      &95      &99      &100      &100     \\
		& & ASR &100    &98    &100    &98    &87    & \multicolumn{1}{c|}{90}   &90      &92      &100      &99      &99      &94      &100    &100    &100    &92    &100    & \multicolumn{1}{c|}{95}   &100      &100      &100      &100      &100      &100     \\ \midrule
		\multirow{2}{*}{\textbf{CSI}}  & &ACC &100    &100    &100    &100    &95    & \multicolumn{1}{c|}{93}   &100      &100      &100      &100      &100      &100      &100    &100    &100    &100    &100    & \multicolumn{1}{c|}{95}   &100      &100      &100      &100      &100      &100     \\
		& & ASR &100    &100    &100    &100    &94    & \multicolumn{1}{c|}{100}   &100      &100      &100      &100      &100      &99      &100    &100    &100    &99    &100    & \multicolumn{1}{c|}{95}   &100      &100      &100      &100      &100      &100     \\ \midrule
		\multirow{2}{*}{\textbf{OSI}}  & &ACC &100    &100    &100    &100    &94    & \multicolumn{1}{c|}{93}   &99      &90      &100      &93      &100      &100      &100    &100    &100    &100    &100    & \multicolumn{1}{c|}{95}   &100      &100      &95      &99      &100      &100     \\
		& & ASR &100    &98    &100    &98    &87    & \multicolumn{1}{c|}{90}   &90      &92      &100      &99      &99      &94      &100    &100    &100    &92    &100    & \multicolumn{1}{c|}{95}   &100      &100      &100      &100      &100      &100     \\ 
		
		\bottomrule
	\end{tabular}
	\label{tab:model}
\end{table*}

\textbf{Impact of the poison durations of \bd.} 
Considering that a typical utterance duration used for enrollment is between 3 to 120 seconds~\cite{zeinali2019short}, we set the default duration to 4s in the following experiments. To evaluate the impact of \bd's contaminated rate on the legitimate user's voice samples, we set the duration of \bd to vary from 1.5s$\sim$4s at an interval of 0.5s, which ensures \bd can be applicable to almost scenarios. We superimpose the backdoor on unseen user's voice samples to form the poisoned enrolling voiceprint, respectively. Based on ECAPA-TDNN, we infer all voiceprints and calculate the similarity scores between different user-poisoned voiceprint pairs and \bd-poisoned voiceprint pairs. Fig.~\ref{fig:duration} shows that \bd can well balance the poisoning rate and the usability of legitimate users with different attack duration because the ACCs are always 100\%. ASR results also demonstrate that a longer duration facilitates \bd to poison enrolling voiceprints, as the 4s duration makes ASRs of the attack 100\% across three scenarios.

\begin{figure}[t]
	\centering
	\subfigure[Librispeech]{
		\includegraphics[width=0.42\textwidth]{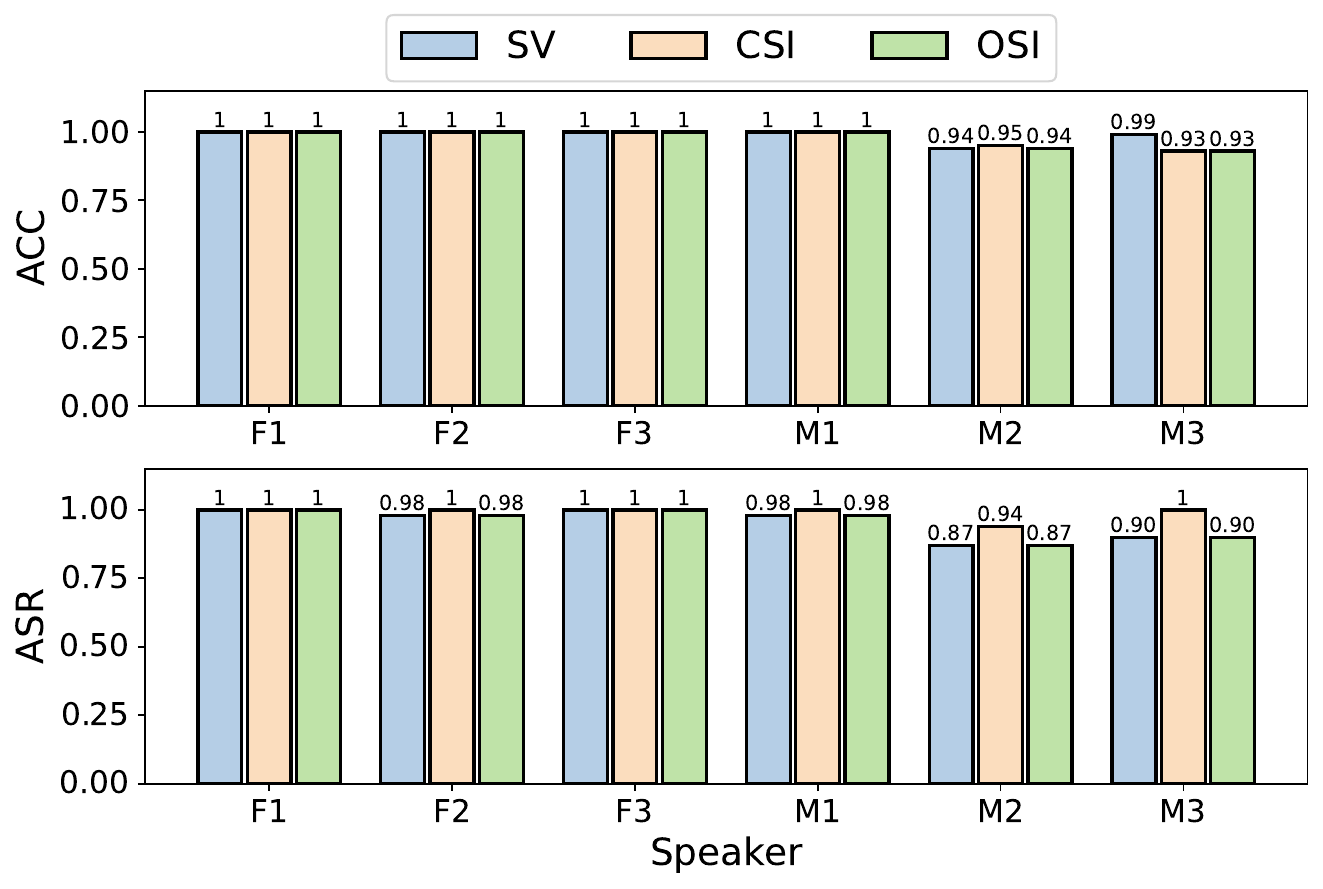}
	}
	\quad
	\hfill 	
	\subfigure[Voxceleb1]{
		\includegraphics[width=0.42\textwidth]{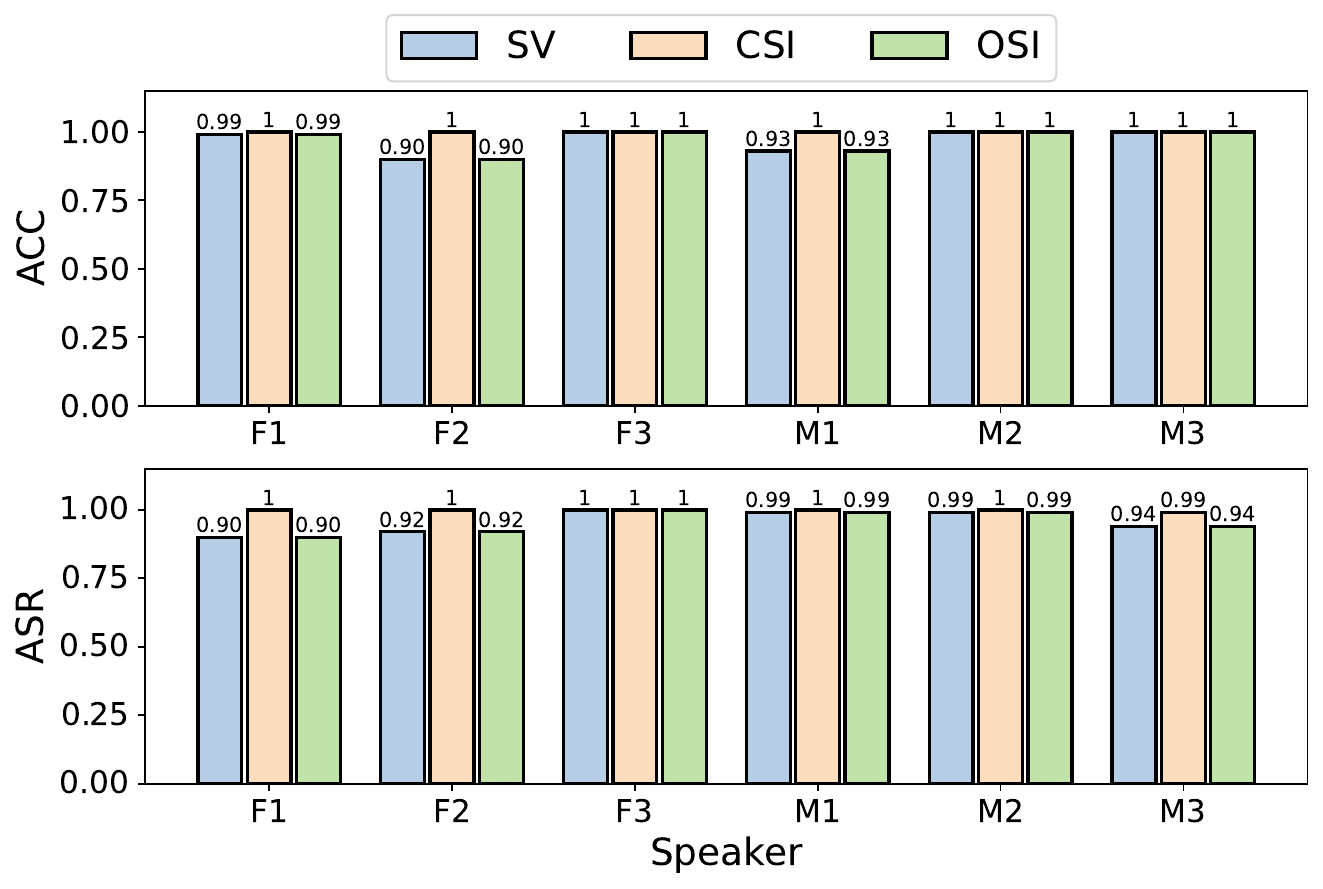}
	}
	\caption{Performance with different speakers (3 males and 3 females) from LibriSpeech and VoxCeleb1, respectively.}
	\label{fig:speaker}
\end{figure}

\textbf{Impact of victim speakers.} The voiceprints of various speakers can differ significantly, while the same speaker may use different utterances for enrollment each time. Thus, we separately choose audio examples of 6 different people (i.e., 3 females and 3 males) from LibriSpeech and Voxceleb1 datasets to evaluate the speaker's impact on \bd. Due to page limitations, the specific speakers are listed in Tab.~\ref{tab:people}. For each speaker, we randomly select 3 utterances for creating the backdoor trigger and another 20 samples for test. Fig.~\ref{fig:speaker} presents the resulting \emph{ACC} and \emph{ASR} for each speaker. Results show that \bd can be effective on both male and female speakers, showing both high confidence in the attackers (i.e., average 95.9\%, 99.8\%, and 95.9\% ASRs) and the legitimate users (i.e., average 97.2\%, 100\%, and 97.2\% ACCs) to bypass the SRS recognition under SV, CSI, and OSI tasks. Although the results of different speakers may be slightly discrepant, our experiment also implies that the speech content does not significantly impact \bd, as the mainstream SRS models are text-independent.

\begin{figure}[t]
	\centering
         \includegraphics[width=0.42\textwidth]{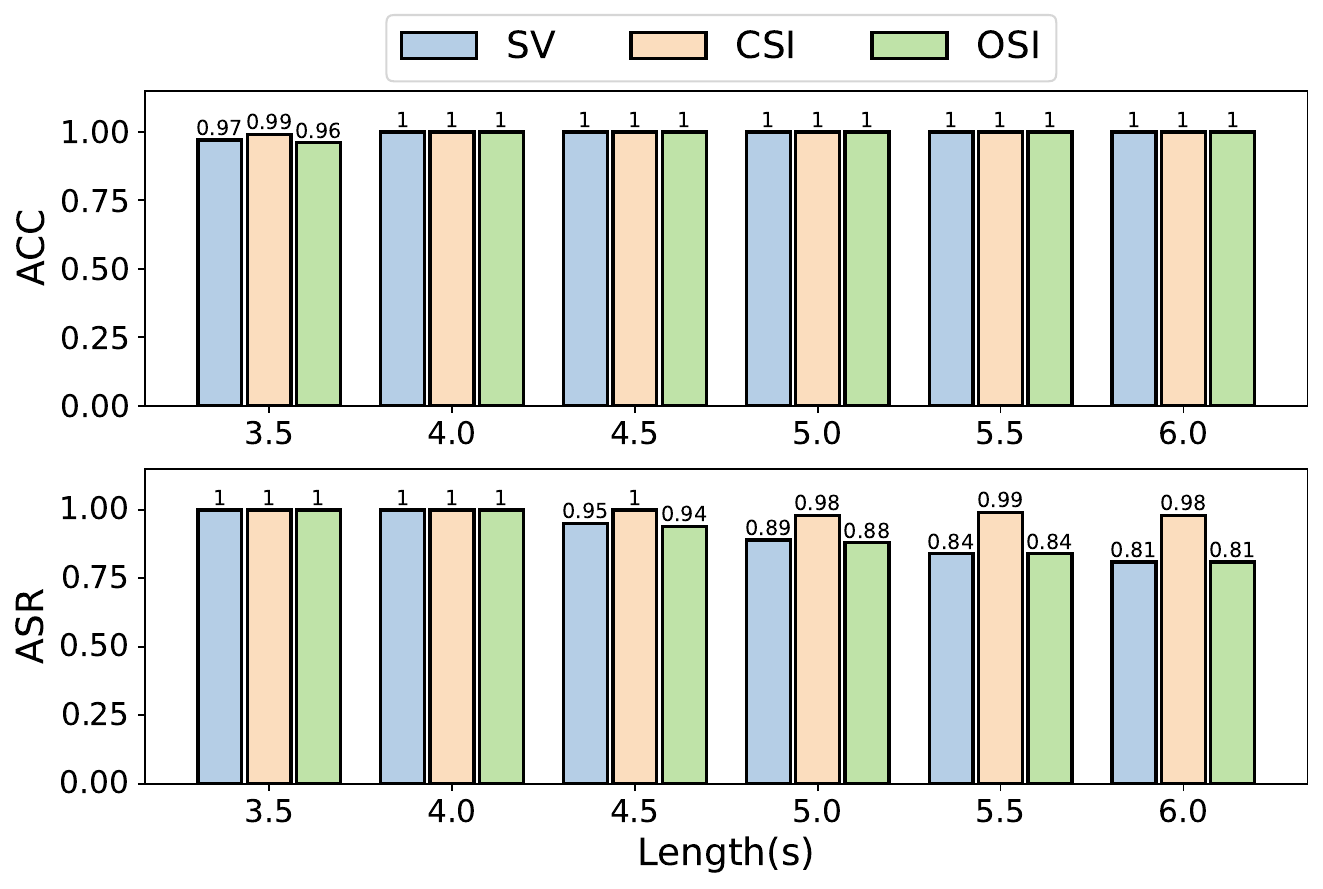}
	\caption{ACC/ASR performance with different victim sample lengths.}
	\label{fig:content_length} 
\end{figure}

\begin{table}[t]
\scriptsize
\centering
\renewcommand\arraystretch{1} 
\setlength\tabcolsep{1.5pt}
\setlength{\abovecaptionskip}{0pt}%
\setlength{\belowcaptionskip}{0pt}%
\caption{Pyannote's thresholds and EERs on different languages}
\begin{tabular}{c|c|c|c|c|c}
\toprule
\textbf{Pyannote}& \multicolumn{1}{c|}{\textbf{Librispeech}} & \multicolumn{1}{c|}{\textbf{AISHELL-1}} & \multicolumn{1}{c|}{\textbf{MLS Italian}} & \multicolumn{1}{c|}{\textbf{MLS German}} & \multicolumn{1}{c}{\textbf{MLS French}} \\ \midrule
\textbf{Threshold} & 0.724 & 0.612 & 0.666 & 0.555 & 0.533 
\\ \midrule
\textbf{EER(\%)}   & 3.53  & 2.96 &  5.51  & 0.64  & 1.68
\\ \bottomrule                                        
\end{tabular}
\label{tab:pyannote_threshold}
\end{table}

\begin{figure}[t]
	\centering
         \includegraphics[width=0.42\textwidth]{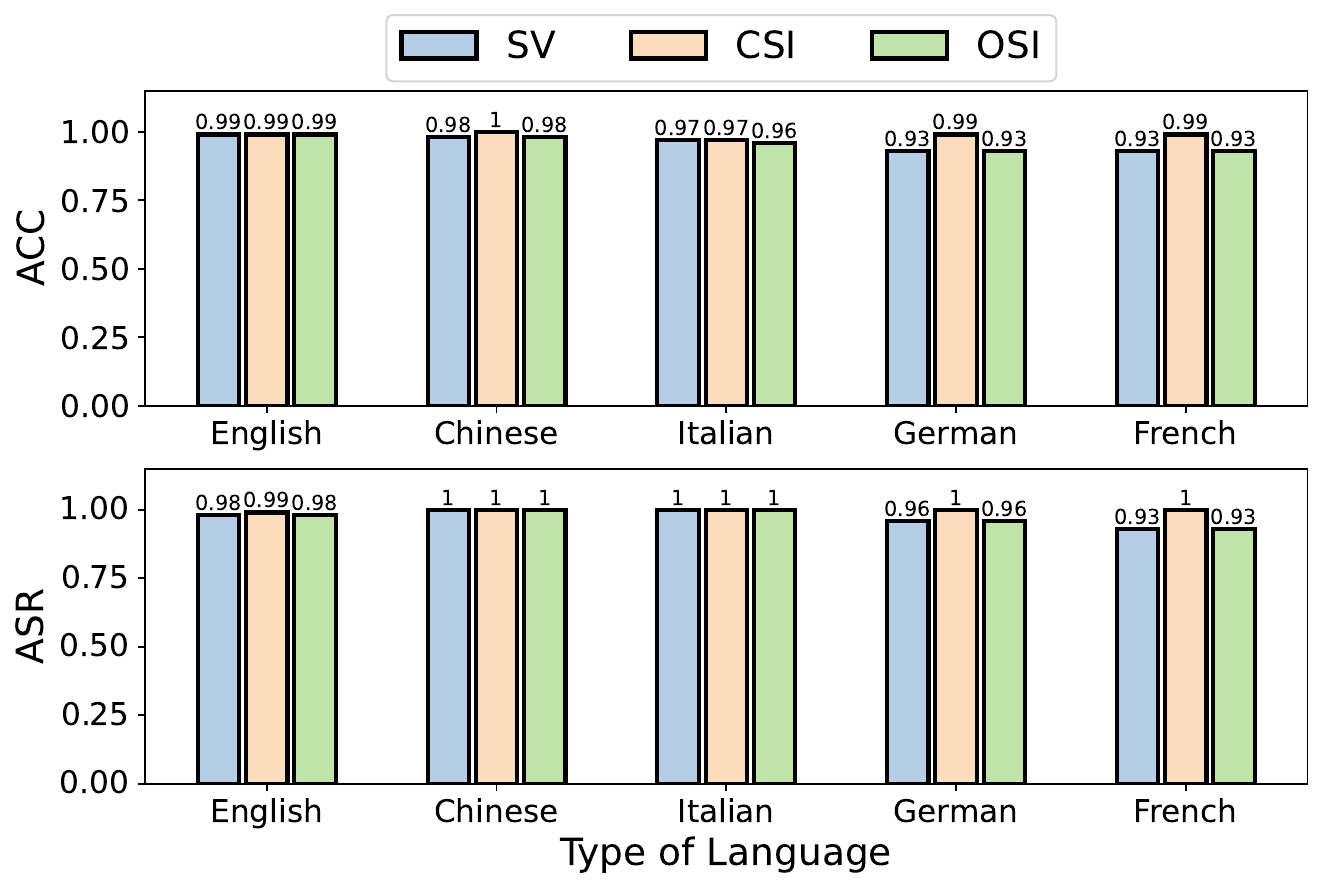}
	\caption{\blue{Performance of \bd facing with different languages.}}
	\label{fig:different_language} 
\end{figure}

\blue{\textbf{Impact of different languages.} 
Upon adopting the \bd framework, adversaries may face situations where victims speak different languages. To evaluate the influence of various languages on \bd, we choose five prominent languages: English, Chinese, Italian, German, and French, along with their representative corpora, as detailed in Tab.~\ref{tab:people}.
Since cross-language issues have been shown to degrade SRS model performance~\cite{thienpondt2020cross}, we fine-tune the Pyannote model for each language, adapting it from English to the other languages separately. Consequently, the corresponding EERs and thresholds for each fine-tuned model are tabulated in Tab.~\ref{tab:pyannote_threshold}.
We randomly select six speakers from each dataset and tailor backdoor triggers for every victim speaker. The averaged performance for each dataset is presented in Fig.~\ref{fig:different_language}. 
Remarkably, the results demonstrate that \bd maintains excellent attack performance across languages, with the minimum ACC and ASR still exceeding 92.5\% and 93.3\%, respectively. In SV/CSI/OSI scenarios, the average ACC and ASR reach up to 97.4\% and 97.7\%, respectively.
This success can be attributed to the fact that the \bd framework crafts a specific backdoor trigger for each victim, enabling the optimization of a poisoned voiceprint (i.e., speaker embedding) that closely resembles both the victim and the trigger in a language-agnostic manner, as long as the SRS model represents speaker embedding normally.
}

\textbf{Impact of victim utterance lengths.} Besides the duration of \bd, we also study the impact of victim samples' length, which characterizes one of the main features of their content. Specifically, we select 20 voice samples for each length, which ranges from 0.5s$\sim$6s at an interval of 0.5s. We maintain \bd's duration as 3.5s. Fig.~\ref{fig:content_length} shows the resulting \emph{ACC} and \emph{ASR} of \bd on each length level of victim samples. \bd barely reduces the accuracy (ACC) of the SRS model for recognizing legitimate users, as almost all ACCs hit 100\% under three tasks, suggesting \bd is not only physically imperceptible (inaudible) but also imperceptible to user experience. We observe the ASRs decrease slightly with the victim sample length increasing. Since the mainstream SRS models adopt average pooling for mapping speech inputs with arbitrary frames to a fixed dimensional voiceprint, we believe that when the victim voice sample gets longer, \bd's poisoning significance for entire speech frames degrades due to such average pooling operation. However, the ASR is still up to 81\% even if the victim's voice (6.0s) is much longer than \bd (3.5s). 

\textbf{Different target SRS models.} We validate the effectiveness of \bd in manipulating the same set of legitimate users in three tasks, which is separately against two representative SRS models (i.e., ECAPA-TDNN~\cite{2020ecapatdnn} and Pyannote~\cite{bredin2020pyannote}). In this experiment, we follow the default black-box configuration and then present the resulting ACC and ASR for each SRS model in Tab.~\ref{tab:model}. Results reveal that \bd can apply to both two SRS models, the average ASRs ranging from 95.6\% (for SV) to 99.4\% (for CSI) on the ECAPA-TDNN model as well as from 98.9\% (for SV\& OSI) to 99.6\% (for CSI) on the Pyannote model. From the ACC perspective, we find that the accuracy of matching legitimate users with contaminated voiceprints is at least 97.1\% for the three tasks against ECAPA-TDNN. Similarly, the ACC is at least 98.9\% for the Pyannote model. We envision the reason for \bd deriving a bit better performance on Pyannote as follows: ECAPA-TDNN performs better on speaker discrimination (lower EER than Pyannote's), in whose feature space the poisoned voiceprint faces slightly more difficulty in getting close to both \bd and victim's voiceprint. 

\begin{table}[t]
\scriptsize
	\centering
	\renewcommand\arraystretch{0.8} 
	\setlength\tabcolsep{1.0pt}
	\caption{Transferability of different speaker recognition models.}
\setlength{\abovecaptionskip}{0pt}%
\setlength{\belowcaptionskip}{0pt}%
\begin{tabular}{cc|c|c|c|c|c|c}
\toprule

   \multicolumn{1}{c}{\begin{tabular}[c]{@{}c@{}}\textbf{Target}\\ \textbf{Model}\end{tabular}}  &  \multicolumn{1}{c|}{(\%)}  & \multicolumn{1}{c|}{\begin{tabular}[c]{@{}c@{}}\textbf{Pyannote}\\ (1.5s)\end{tabular}} & \multicolumn{1}{c|}{\begin{tabular}[c]{@{}c@{}}\textbf{U-Level}\\ (2.0s)\end{tabular}} & \multicolumn{1}{c|}{\begin{tabular}[c]{@{}c@{}}\textbf{WavLM-Xvec}\\ (4.0s)\end{tabular}} & \multicolumn{1}{c|}{\begin{tabular}[c]{@{}c@{}}\textbf{SpeakerNet}\\ (3.5s)\end{tabular}} & \multicolumn{1}{c|}{\begin{tabular}[c]{@{}c@{}}\textbf{D-vector}\\ (2.5s)\end{tabular}} & \multicolumn{1}{c}{\begin{tabular}[c]{@{}c@{}}\textbf{ResNet34}\\ (2.0s)\end{tabular}} \\ \midrule
\multicolumn{1}{c}{\multirow{2}{*}{\textbf{SV}}} & ACC & 99     & 99       & 96    & 89         & 98   & 99       \\ 
\multicolumn{1}{c}{}                    & ASR & 99     & 98       & 41    & 71         & 97   & 98       \\ \midrule
\multirow{2}{*}{\textbf{CSI}}                    & ACC & 100    & 100      & 95    & 77         & 100  & 100      \\ 
                                        & ASR & 100    & 100      & 79    & 99         & 100  & 100      \\ \midrule 
\multirow{2}{*}{\textbf{OSI}}                    & ACC & 99     & 99       & 95    & 73         & 98   & 99       \\ 
                                        & ASR & 98     & 98       & 40    & 71         & 97   & 98       \\ 
\bottomrule                                        
\end{tabular}
\label{tab:transferability}
\end{table}

\textbf{Transferability of \bd.} Given that an adversary can hardly have prior knowledge of the exact parameters and structures of SRS models, and especially commercial SRSs' APIs are protected with limited query counts, casting challenges in manipulating the system via gradient estimation. In this regard, we consider a more practical attack model, i.e., crafting backdoors based on the open-source ECAPA-TDNN. Then these examples directly attack other unseen black-box SRSs under the default configuration. Specifically, we employ Pyannote, U-Level, WavLM-Xvec, SpeakerNet, D-vector, and ResNet34 to examine the transferability of \bd. 
Tab.~\ref{tab:transferability} lists the performance of \bd attacking different black-box models under three tasks, indicating that \bd features significant transferability and can achieve more than 95\% ACC/ASR on most unseen models.
We also observe that the appropriate duration of \bd (i.e., poisoning rate) varies with different models. For instance, \bd can attack Pyannote well, bringing ACCs and ASRs close to 100\% when generated with an 1.5s-duration setting against ECAPA-TDNN. Differently, the poisoning rate needs to be maximized (i.e., 4s as the victim utterance's length) to ensure a higher ASR when attacking WavLM-Xvec.
Notably, WavLM-Xvec has a significantly larger parameter size of up to 316.62M compared to ECAPA-TDNN's 6.2M. However, we believe that such large models may not be as practical for deployment on lightweight smart devices intended for access control purposes.

\subsection{Physical Attack Performance}
\begin{figure}[t]
	\centering 
	\includegraphics [width=0.3\textwidth]{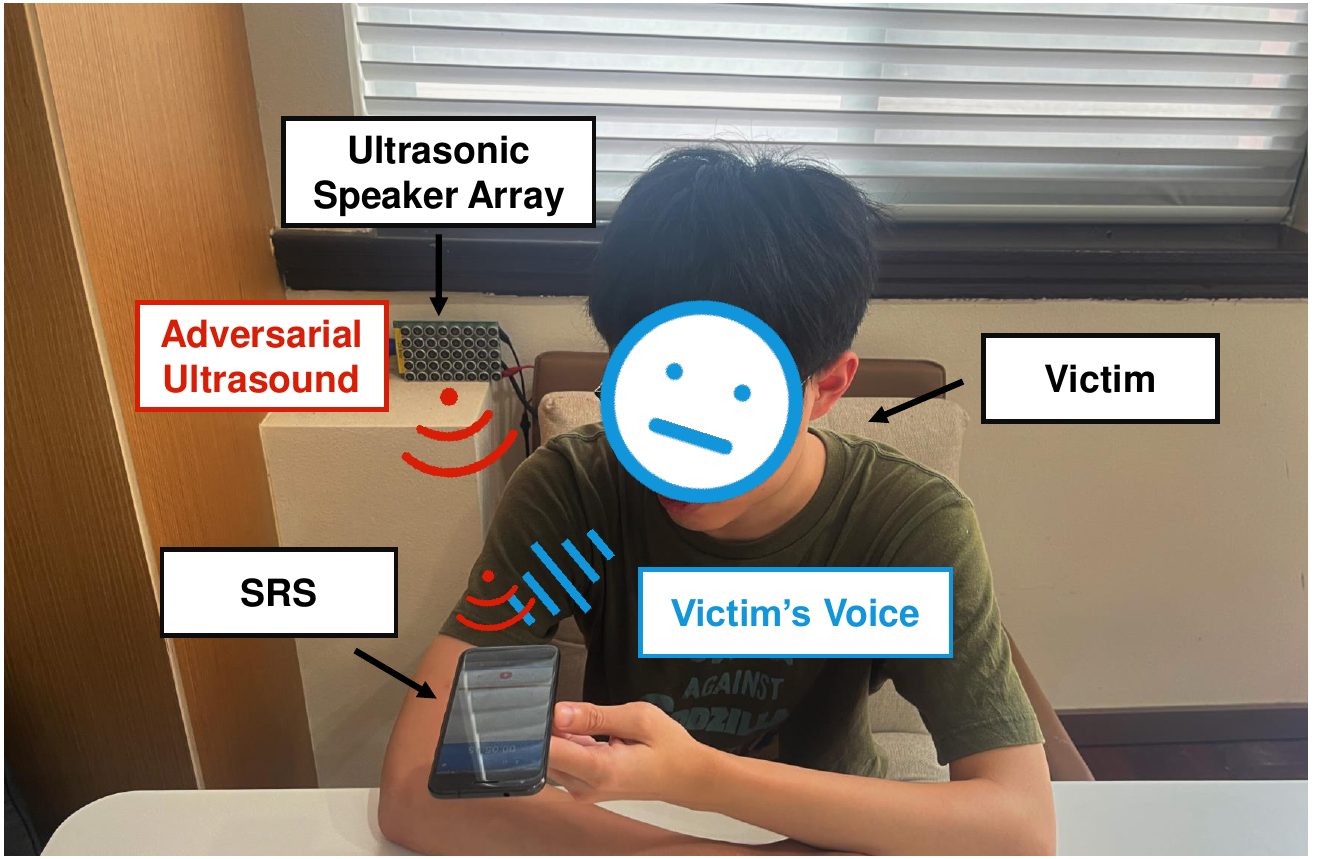}
	\caption{The experimental setup of \bd in the physical world scenario.}
	\label{fig:setup}
\end{figure}

We carry out extensive experiments in the physical domain to evaluate the practical performance of \bd under different conditions, i.e., environments, distances, angles, and recording devices.
Fig.~\ref{fig:setup} presents our experimental setup where the adversarial ultrasound is emitted by a customized ultrasonic speaker array while the victim smartphones (Google Pixel, OPPO Reno5, iPhoneX, and Samsung S6) are utilized as receivers. \blue{We conduct physical evaluations in relatively quiet environments (36$\sim$40dB, with slight HVAC noises) using Pixel as the default receiver, because users tend to enroll their voiceprints in a scene without obvious noise interference. Notably, we also perform the noise-related experiment to assess \bd comprehensively.}

\begin{figure}[t]
	\centering
	\includegraphics[width=0.45\textwidth]{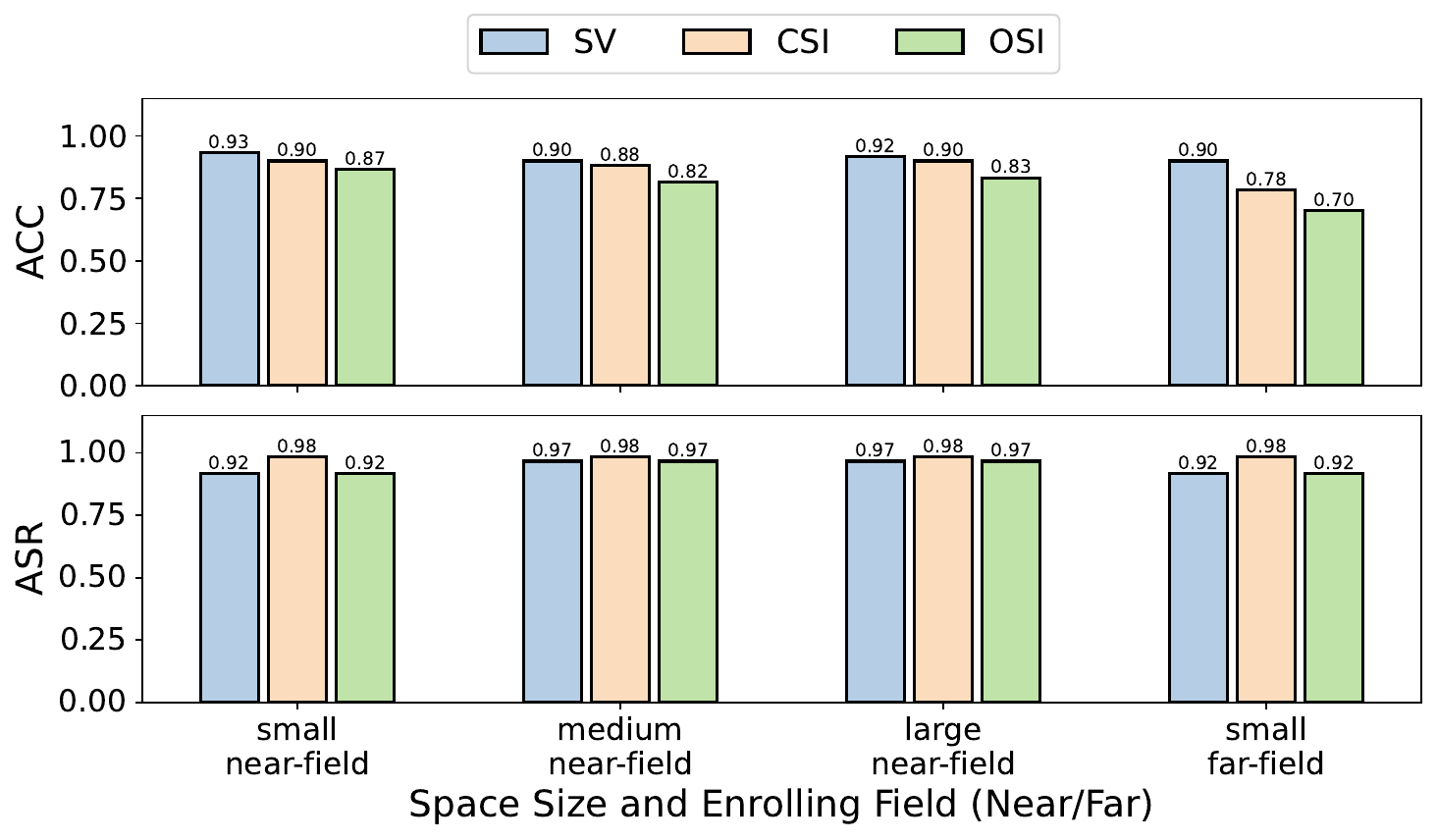}
	\caption{ACC/ASR of \bd attacking in different rooms (varying space sizes and the users performs near- \& far-field enrollment).}
	\label{fig:reverb} 
\end{figure}

\textbf{Attacking when the victims enroll in different environments.} Our experiments are conducted in three space sizes---small office (2.4m$\times$2.6m, 36dB), medium lounge (6.3m$\times$3.8m, 40dB), and large meeting room (12m$\times$6.4m, 38dB). In these conditions, the reverberation pattern of audible sound waves varies with space size.
Our configuration considers a typical user enrollment scenario, where the user-receiver distances are generally within 0.3m (i.e., near-field) in quiet surroundings when users perform voiceprint registration. The ultrasonic speaker array is 1m away from the recording device, respectively. 
Fig.~\ref{fig:reverb} shows that \bd maintains high ASRs (all$\ge$$91.7\%$) in all rooms and appears slightly better in a larger space. Besides, the increasing space might lower ACCs, which we believe is due to the audible sound reflections from the walls getting weaker when the room size increases. However, it is worth noting that the inaudible voice attacks are barely affected by reverberation due to their high-frequency signal direct injection into the microphone, with weak acoustic diffraction. Namely, the energy of recorded human voice is mainly contributed by the direct path, making \bd's poisoning effect more pronounced. 

In addition, we investigate an uncommon distance of users enrolling at 2m from the recording device (i.e., far-field, a corner case) in the small office. We obtain the far-field ACC: 90\%, 78.3\%, 70\% and ASR: 91.7\%, 98.3\%, 91.7\%. Compared with the near-field ACC/ASR, it suggests far-field recorded voice samples indeed introduce challenges to speaker recognition tasks as revealed by prior works~\cite{jin2007far,gusev2020deep} due to multiple challenges, such as low audio fidelity and complex reverberation.

\begin{figure*}[t]
	\centering  
	\subfigure[\blue{Impact of Noise Type}]{ 
		\begin{minipage}[t]{0.31\textwidth}
			\raggedright
			\includegraphics[width=1.05\textwidth]{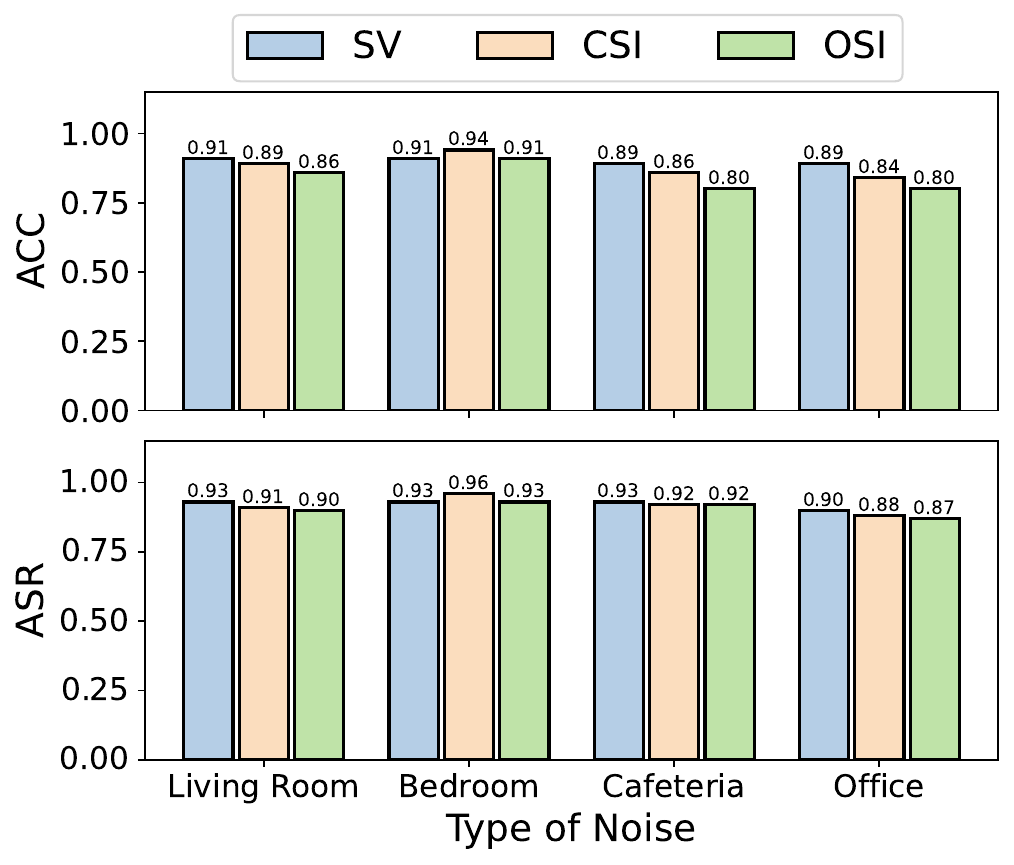}
		\end{minipage}
	}
	\subfigure[{Impact of Angle}]{   
		\begin{minipage}[t]{0.31\textwidth}
			\centering
			\includegraphics[width=1.065\textwidth]{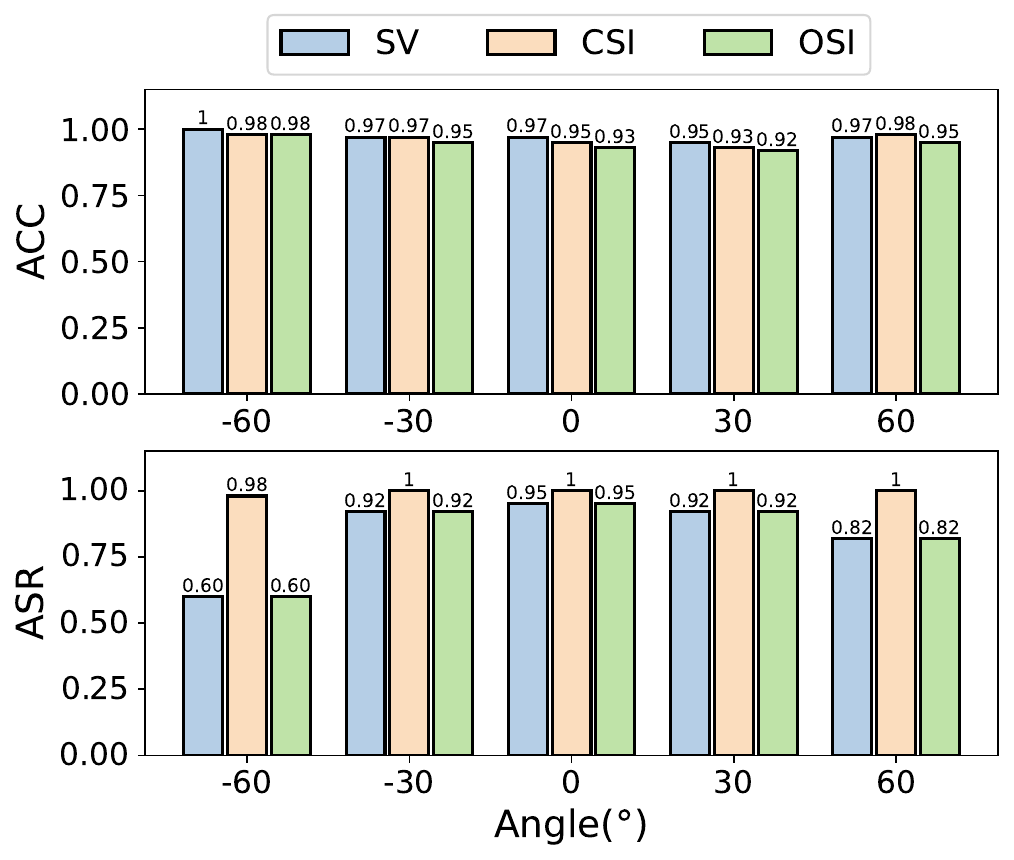}
		\end{minipage}
	}
	\subfigure[{Impact of Recording Device}]{ 
		\begin{minipage}[t]{0.31\textwidth}
			\raggedleft
			\includegraphics[width=1.06\textwidth]{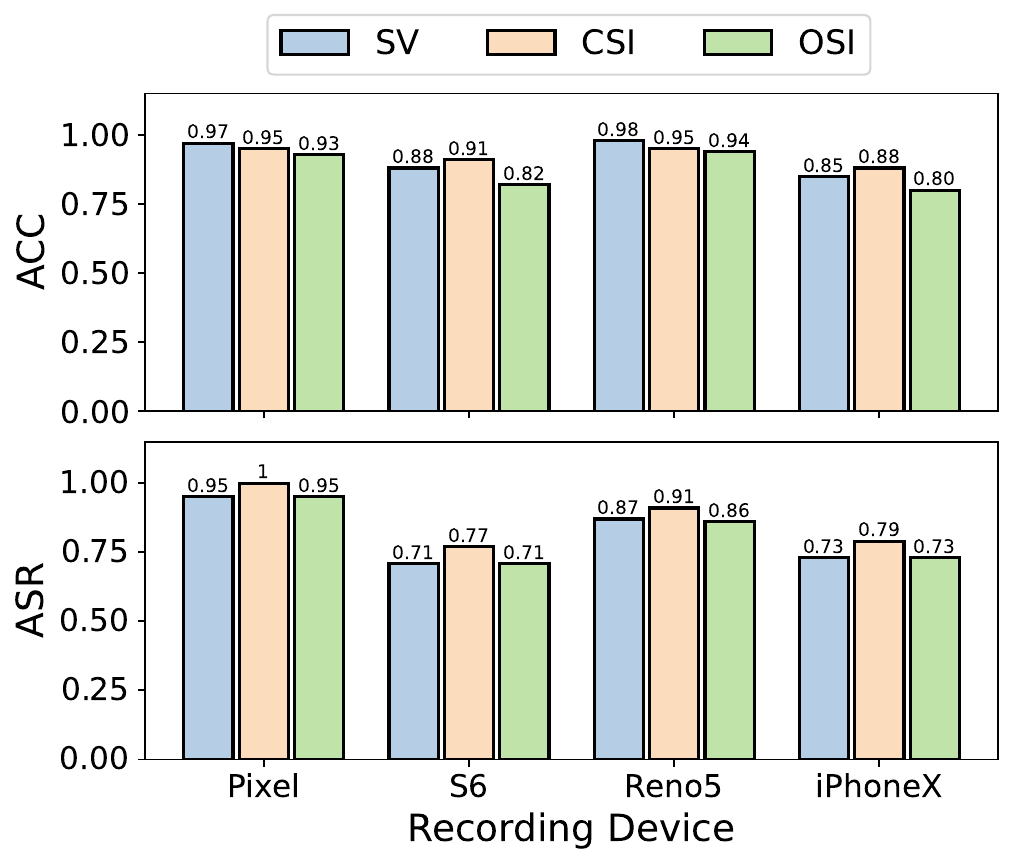}
		\end{minipage}
	}
	\vspace{-5pt}
	\caption{\blue{The impacts of (a) noises, (b) angles, and (c) recording devices on ACC/ASR performance of \bd.}}  
	\label{fig:eval_physical}    
\end{figure*}

\blue{\textbf{Impact of attack distances.}
Apart from the different environments where users may register voiceprints, we also consider how the performance of \bd is affected by the distance between the ultrasonic transmitter and the recording device. 
To enable distance-adaptive attacks that encompass a wide range of daily scenarios (e.g., using smart speakers from 3m away), we incorporate a power amplifier~\cite{amplifierHSA4015} and vary the attack distance from 50$\sim$600~cm.
Fig.~\ref{fig:eval_physical}(a) shows the \emph{ACC} and \emph{ASR} corresponding to each distance, where the ACCs are consistently high (all$\ge$92\%). Notably, the ASRs demonstrate that \bd performs effectively (achieving $\ge$96\% ASR in the CSI setting) even at 400~cm. This success is attributed to our consideration of ultrasound attenuation, relative power augmentation during the optimization, and the employment of an amplifier to maintain efficient backdoor triggers at different distances. Moreover, we observe a decrease in ASR with increasing attack distance, while the ACCs increase with the attack distances, indicating a reduction in the poisoning effect of the triggers due to relative energy mismatch. 
We find that even attacking at 600~cm away, which covers most everyday scenarios, \bd can still potentially work. However, we cannot boost the emission power infinitely as the airborne demodulation will leak the audible sound once a certain power is reached~\cite{iijima2018audio}.
We also discover that specifically tuning the hyper-parameter $\beta$ to $[1.5,3]$, which mimics a relatively louder user voice and a weaker \bd volume for such a far-distance attack, can significantly improve its ASR performance to 89\%.}

\begin{figure}[t]
	\centering
	\includegraphics[width=0.45\textwidth]{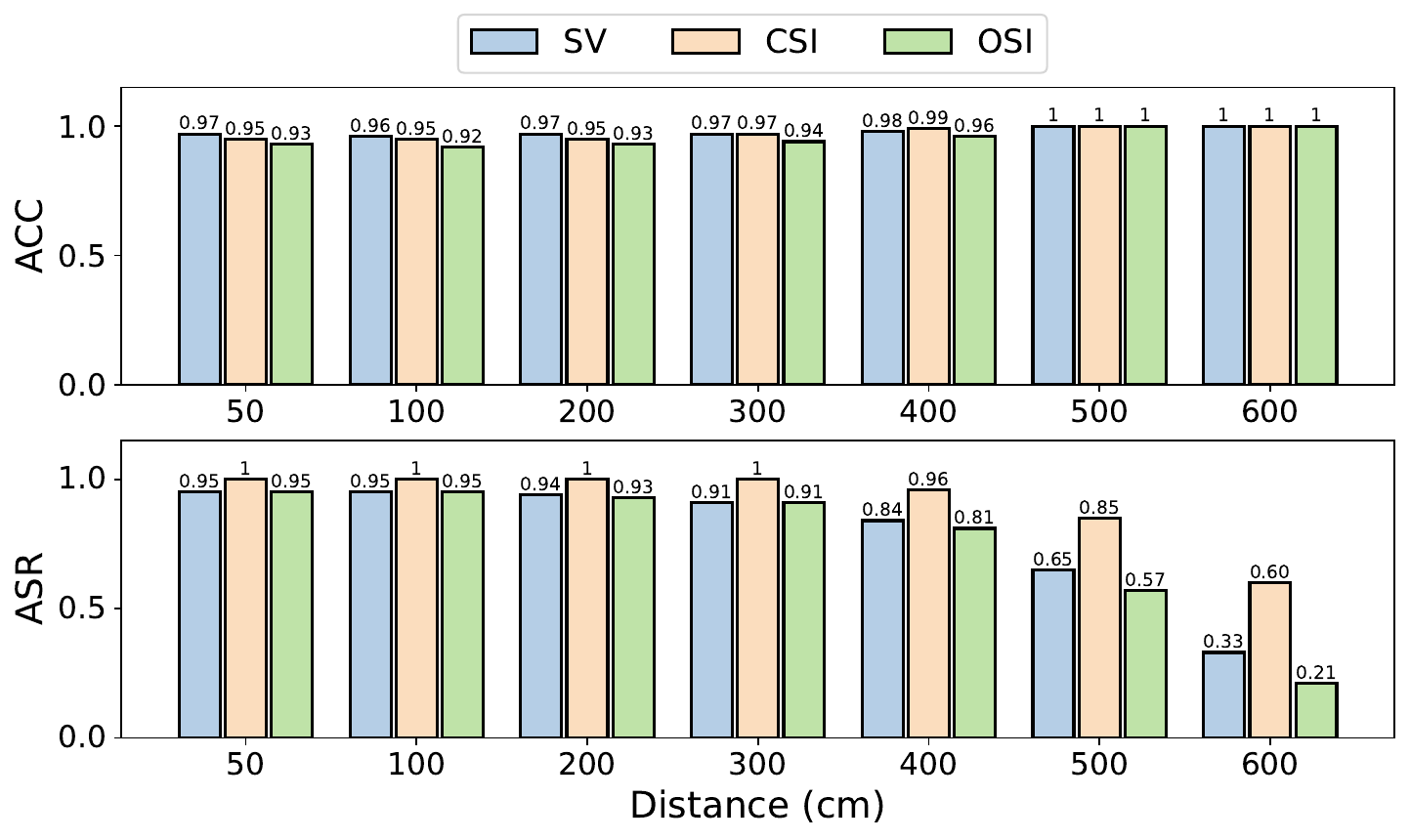}
	\caption{\blue{The performance of \bd at different attack distances.}}
	\label{fig:noises} 
\end{figure}

\blue{\textbf{Impact of ambient noises.}
Users typically interact with SRSs in noise-free environments. However, despite the well-documented negative effect of noise on SRSs~\cite{mandasari2012effect}, users may still opt to enroll their voiceprints if the interference remains minimal.
For instance, SRSs can maintain acceptable performance with a signal-to-noise ratio (SNR) of 20dB~\cite{mandasari2012effect}.
To assess the effect of various noises on the \bd system, we select four typical environmental scenes, namely living room (TV show), bedroom (fan running), cafeteria (people chatting), and office (keyboard typing). These audio samples obtained from the Freesound database~\cite{freesound}, are played continuously while simultaneously playing victim speech samples and launching triggers.
Fig.\ref{fig:noises} demonstrates that the noises slightly degrade \bd's attack performance compared to the noise-free baseline, where ACCs and ASRs are nearly 100\%. Among the scenarios, the bedroom fan running has the least impact on \bd with ACCs/ASRs still exceeding 91\%/93\%. 
However, the office noise case experiences a more significant performance drop with 80\% ACC and 87\% ASR under the OSI setting due to the intense high-frequency energy of crisp noises caused by keyboard typing and mouse striking. This reduction in performance may be attributed to \bd's primary influence on low-frequency (0-4kHz) acoustic features through demodulation, making it susceptible to high-frequency noise that can diminish its attack performance on deceiving SRS models.
}

\textbf{Impact of injection angles.} We examine the different injection angles of the ultrasonic transmitter at 50~cm away from the recording device. The test angle changes (-60$^{\circ}$$\sim$60$^{\circ}$) with a step of 30$^{\circ}$, where 0$^{\circ}$ indicates that the ultrasonic speaker aims directly at the recording device's bottom microphone. As shown in Fig.~\ref{fig:eval_physical}(b), we can observe that \bd can achieve relatively high ASRs at 0$^{\circ}$, and the ACC hardly changes (all$\ge$ 92\%) because the human voice have large wavelengths and propagate uniformly in the sound field. Although the off-the-shelf microphones deployed on smart devices are omnidirectional, the effective energy of the ultrasound carrier injected into the microphone reduces  with the injection angle turning larger, thus resulting in ASR to decrease.



\textbf{Impact of recording devices.} We further investigate whether \bd works well on different recording devices, including Google Pixel, OPPO Reno5, iPhone X, and Samsung S6. Fig.~\ref{fig:eval_physical}(c) shows that tuner is able to achieve an average 95\% ACC and 96.7\% ASR on Pixel, as well as 95.7\% ACC and 87.8\% ASR on Reno5. 
In sum, we observe that devices with higher ACCs/ASRs usually feature better transmitter-receiver frequency response $H_t(f)$ proposed in Sec.~\ref{Signal Transmission}, causing higher SNR of the demodulated baseband.
This corresponds to that the poisoning effect of \bd relatively decrease (i.e., with lower ASRs) on S6 and iPhoneX, we believe it is due to the best carrier frequency of the ultrasound transmitter varying with the recording device, which is reported in \cite{zhang2017dolphinattack}. We envisage that applying ultrasonic speaker with adjustable frequencies (e.g., 26.3kHz) can improve \bd's performance on different recording devices.

\color{black}
\subsection{Physical Deployment Considerations}
The experiment results above validate that \bd can contaminate voiceprint under various impact factors while ensuring that the victim remains unaware of the presence of attacks. Moreover, \bd does not necessitate access to the target SRS system's training data. When deploying \bd prototype in a practical scenario, two aspects need to be considered.


One aspect pertains to the uncommon property of the custom ultrasonic transmitter, which could raise suspicion in everyday scenarios, potentially alerting the victim despite \bd operating in complete inaudibility during the attack. 
We believe that such a concern can be mitigated by placing the ultrasonic transmitter at a distance, e.g., 600~cm as demonstrated in our experiments. The long-range advantage of \bd facilitates its operation through windows or doors. In contrast, traditional AEs/backdoors that utilize loudspeakers are required to be close to the victim (e.g., 1.5~m reported in \cite{yuan2018commandersong}). Moreover, we anticipate that \bd can be designed to be portable and miniaturized as has been demonstrated in the literature~\cite{li2024vrifle}.

The other aspect is related to the nature of ultrasound. Namely, ultrasound encounters difficulties in reaching recording devices in non-line-of-sight (NLoS) scenarios due to its poor diffraction and high attenuation when obstructed by obstacles. This drawback presents a challenge for all ultrasound-based injection methods~\cite{zhang2017dolphinattack,roy2018inaudible} in NLoS cases primarily because of their inherent high frequency. Fortunately, we envision that combining SurfingAttack~\cite{yan2020surfingattack} can bridge this gap, even if there are obstacles on the table.

\color{black}
\subsection{Robustness to Defenses}
\textbf{Against audio pre-processing defense methods.} Previous works~\cite{deng2022fencesitter,li2020advpulse,yuan2018commandersong} render commonly used defenses against audio adversarial example attacks. Since \bd is crafted by adversarial training against the given SRS model, we are driven to examine its robustness under 6 representative defenses adopted by prior studies, including voice activity detection (VAD), quantization, MP3 compression, band-pass filter, median filter, and squeezing. We evaluate all three tasks on the Pyannote model and consider 1) \emph{naive attacker} and 2) \emph{adaptive attacker}, respectively. In the benign user enrollment case (without attack), the ACC is 100\%, 100\%, 100\%, and the ASR is 0\%, 8.6\%, 0\%. In contrast, in adversarial user enrollment (with attack), ACC is 99.2\%, 99.2\%, 99.2\%. ASR is 98.8\%, 99.2\%, 98.8\%.

The results of \emph{naive attacker} scenarios are given in four Tab.~\ref{tab:definitive},\ref{tab:bandpass},\ref{tab:squeezing},\ref{tab:median}. The attack performance against three representative defense methods and the definitive configurations are shown in Tab.~\ref{tab:definitive}, where the VAD (i.e., normalized threshold: -25dB according to the audio's maximum), MP3 compression (i.e., from ``WAV'' to ``MP3'' format), and quantization (i.e., conversion from 16bit to 8bit) barely reduce the ASRs (all$\ge$97.6\%). We also conduct fine-grain experiments on three other typical defenses. Tab.~\ref{tab:bandpass},\ref{tab:squeezing} demonstrate that \bd can resist the band-pass filter and squeezing\footnote{\blue{Squeezing rate: e.g., down-sampling 16,000 Hz to 8,000 Hz and then the missing information is up-sampled to 16,000 Hz when the rate is 0.5.}} under different settings (most ASRs $\ge$93.8\%). \bd's performance decreases when facing the median filter as shown in the left half of Tab.~\ref{tab:median}, especially ASR is down to 65.8\% when the kernel is 5.

\begin{table}[t]
\scriptsize
\centering
\renewcommand\arraystretch{0.8} 
\setlength\tabcolsep{2.8pt}
\setlength{\abovecaptionskip}{0pt}%
\setlength{\belowcaptionskip}{0pt}%
\caption{Definitive Defenses}
\label{tab:definitive}
\begin{tabular}{cc|c|c|c|c}
\toprule
\multicolumn{1}{l}{}           &   (\%)  & \textbf{No modification} & \textbf{VAD} & \textbf{MP3 compress} & \textbf{Quantization} \\ \midrule
\multicolumn{1}{c}{\multirow{2}{*}{\textbf{SV}}}                               & ACC & 99.2                                            & 91.1                                    & 99.2                                        & 98.9                                         \\  
\multicolumn{1}{c}{}   & ASR & 98.8                                            & 98.9                                   & 98.1                                        & 98.8                                         \\ \midrule
\multirow{2}{*}{\textbf{CSI}}                               & ACC & 99.2                                            & 96.3                                & 99.2                                        & 99.0                                         \\  
 & ASR & 99.0                                            & 99.2                                & 99.0                                        & 99.0                                         \\ \midrule
\multirow{2}{*}{\textbf{OSI}}                               & ACC & 99.2                                            & 91.1                                & 99.2                                        & 98.9                                         \\ 
& ASR & 98.8                                            & 98.9                                & 98.1                                        & 98.8                                         \\ 
\bottomrule
\end{tabular}
\end{table}

\begin{table}[t]
\scriptsize
\centering
\renewcommand\arraystretch{0.8} 
\setlength\tabcolsep{4.8pt}
\caption{Defense with Band-pass Filter}
\setlength{\abovecaptionskip}{0pt}%
\setlength{\belowcaptionskip}{0pt}%
\label{tab:bandpass}
\begin{tabular}{cc|c|c|c|c|c}
\toprule
\multicolumn{2}{c|}{\textbf{Cut-off Freq (Hz) (\%)}}      & \multicolumn{1}{c}{\textbf{4000}} & \multicolumn{1}{c}{\textbf{5000}} & \multicolumn{1}{c}{\textbf{6000}} & \multicolumn{1}{c}{\textbf{7000}} & \multicolumn{1}{c}{\textbf{8000}} \\ \midrule
\multicolumn{1}{c}{\multirow{2}{*}{\textbf{SV}}}& ACC & 98.9                                 & 98.9                                 & 98.9                                 & 98.9                                 & 98.9                                 \\
  & ASR & 95.9                                 & 93.8                                 & 94.1                                 & 91.5                                 & 97.6                                 \\ \midrule
                               & ACC & 99.6                                 & 99.6                                 & 99.6                                 & 99.6                                 & 98.9                                 \\ 
\multirow{-2}{*}{\textbf{CSI}} & ASR & 99.0                                 & 99.0                                 & 99.0                                 & 99.0                                 & 99.0                                 \\ \midrule
                               & ACC & 99.5                                 & 99.5                                 & 99.5                                 & 99.5                                 & 98.8                                 \\  
\multirow{-2}{*}{\textbf{OSI}} & ASR & 95.9                                 & 93.8                                 & 94.1                                 & 91.5                                 & 97.6                                 \\ 
\bottomrule                                        
\end{tabular}
\end{table}

\begin{table}[t]
\scriptsize
\centering
\renewcommand\arraystretch{0.8} 
\setlength\tabcolsep{3.8pt}
\setlength{\abovecaptionskip}{0pt}%
\setlength{\belowcaptionskip}{0pt}%
\caption{Defense with Squeezing}
\label{tab:squeezing}
\begin{tabular}{cc|c|c|c|c|c|c|c}
\toprule

\multicolumn{2}{c|}{\textbf{Squeezing Rate (\%)}}      & \multicolumn{1}{c}{\textbf{0.3}} & \multicolumn{1}{c}{\textbf{0.4}} & \multicolumn{1}{c}{\textbf{0.5}} & \multicolumn{1}{c}{\textbf{0.6}} & \multicolumn{1}{c}{\textbf{0.7}} & \multicolumn{1}{c}{\textbf{0.8}} & \multicolumn{1}{c}{\textbf{0.9}}\\ \midrule
\multicolumn{1}{c}{\multirow{2}{*}{\textbf{SV}}}& ACC & 98.3                                & 98.3                                & 98.3                                & 98.3                                & 98.3                                & 98.3        & 98.3        \\
  & ASR & 97.9                                & 89.2                                & 98.4                                & 91.7                                & 96.9                                & 95.8        & 96.9        \\ \midrule
                               & ACC & 98.5                                & 98.5                                & 97.7                                & 98.5                                & 98.5                                & 98.5        & 98.5\\ 
\multirow{-2}{*}{\textbf{CSI}} & ASR & 99.0                                & 98.8                                & 99.0                                & 99.0                                & 99.0                                & 99.0        & 99.0        \\ \midrule
                               & ACC & 97.9                                & 97.9                                & 97.2                                & 97.9                                & 97.9                                & 97.9        & 97.9        \\  
\multirow{-2}{*}{\textbf{OSI}} & ASR & 97.9                                & 89.2                                & 98.4                                & 91.7                                & 96.9                                & 95.8        & 96.9        \\ 
\bottomrule                                        
\end{tabular}
\end{table}

\begin{table}[t]
\scriptsize
\centering
\renewcommand\arraystretch{0.8} 
\setlength\tabcolsep{4.8pt}
\setlength{\abovecaptionskip}{0pt}%
\setlength{\belowcaptionskip}{0pt}%
\caption{Defense with Median Filter}
\label{tab:median}
\begin{tabular}{cc|c|c|c|c|c|c|c|c}
\toprule
\multicolumn{2}{c|}{\multirow{2}{*}{\textbf{Kernel (\%)}}}      & \multicolumn{4}{c|}{\textbf{Naive Adversary}}                                                                                                      & \multicolumn{4}{c}{\textbf{Adaptive Adversary}}   \\ 
                               &              & \multicolumn{1}{c}{\textbf{3}} & \multicolumn{1}{c}{\textbf{5}} & \multicolumn{1}{c}{\textbf{7}} &\multicolumn{1}{c|}{\textbf{9}} & \multicolumn{1}{c}{\textbf{3}} & \multicolumn{1}{c}{\textbf{5}} & \multicolumn{1}{c}{\textbf{7}} & \multicolumn{1}{c}{\textbf{9}} \\ \midrule
\multicolumn{1}{c}{\multirow{2}{*}{\textbf{SV}}}& ACC & 99.2                              & 99.2                              & 99.2                              & 99.2                              & 100      & 99.2      & 99.2      & 99.2      \\
  & ASR & 78.6                              & 65.8                              & 77.6                              & 77.6                              & 93.7      & 84.3      & 90.4      & 90.4      \\ \midrule
                               & ACC & 98.9                              & 99.9                              & 99.9                              & 99.9                              & 98.2      & 99.2      & 99.2      & 99.2      \\ 
\multirow{-2}{*}{\textbf{CSI}} & ASR & 99.0                              & 98.0                              & 99.0                              & 99.0                              & 100      & 99.2      & 99.2      & 99.2      \\ \midrule
                               & \textbf{ACC} & 98.9                              & 99.9                              & 99.9                              & 99.9                              & 98.2      & 99.2      & 99.2      & 99.2      \\  
\multirow{-2}{*}{\textbf{OSI}} & \textbf{ASR} & 78.6                              & 65.8                              & 77.6                              & 77.6                              & 93.7      & 85.0      & 90.4      & 90.4 \\ 
\bottomrule    
\end{tabular}
\end{table}

Moreover, we carry out \emph{adaptive attacker} experiments to evaluate whether combining the median filter into \bd's optimization process can increase its possibility of bypassing such a defense. The right half of Tab.~\ref{tab:median} demonstrates that an attacker can boost the ASR by at least 18.5\% despite the median filter existing.

\textbf{Against the defense for inaudible voice attacks.} \bd essentially performs attacks by modulating a carefully designed backdoor trigger on the ultrasound carrier, i.e., in an inaudible voice attack manner. Therefore, we adopt the representative inaudible voice attack detection method---LipRead~\cite{roy2018inaudible}, which analyzes three aspects: power in sub-50Hz, correlation coefficient, and amplitude skew. We strictly follow the instructions of LipRead and obtain a detection classifier for follow-up evaluation. Then we test LipRead on detecting \bd samples. The success rate of bypassing LipRead is up to 87.9\%. We consider that LipRead cannot detect \bd well due to the following reasons: 1) our backdoors are distinct from the human voice that has various frequencies and are demonstrated to easily concentrate in sub-50Hz~\cite{roy2018inaudible}; 2) \bd can be regarded as the optimized combinations of sine waves, whose sampling points' amplitudes are close to perfect symmetry and do not appear skewness.




\section{Discussion and Future Work}
\color{black}
\subsection{Countermeasures against \bd} 
\textbf{An Unsupervised Detection Method.}
\bd has been demonstrated to resist common signal pre-processing techniques and the classical inaudible voice attack detection methods well. To mitigate this newly discovered threat, we adopt NormDetect~\cite{li2023learning} that has been demonstrated to protect billions of legacy devices instantly in an unsupervised manner. We reproduce NormDetect using 30,042 benign samples from the open-source Fluent Speech Commands. At the same time, we leverage dual-channel information to enhance NormDetect. Specifically, it not only reconstructs each channel's audio spectrum and obtains the anomaly score, where the larger score indicating that it is tend to be attack, but also calculates the spectrum similarity between two channels. We fine-tune the hyper-parameter, i.e., weights of  Our evaluation on the benign and attack audios denotes that NormDetect derives 97.82\% on detecting \bd. We consider it is due to the intrinsic sound field distribution property and nonlinearity effects of ultrasound-based attacks are very significant to be detected.

\textbf{Other Potential Countermeasures.}
In addition to the sophisticated software-based mitigation, we envision that leveraging the inherent differences between ultrasound and audible voice~\cite{he2019canceling,zhang2021eararray}, as well as adopting non-speech-based biometric authentication~\cite{lu2018lippass,yan2019catcher} can be instrumental in enhancing the security of personal identification or access control systems. GuardSignal~\cite{he2019canceling}  draws inspiration from the ANC (active noise canceling) methods might cancel out the ultrasound carrier of our modulated backdoor, thus thwarting the attack due to the weakened demodulated \bd's energy. Note that it requires additional equipment to actively emit ultrasound signals, making integration into the compact smart devices challenging and power-consuming.
On the other hand, EarArray~\cite{zhang2021eararray} operates in a passive manner by leveraging multiple microphones and taking advantage of the ultrasound's unique propagation characteristics, such as high directivity and attenuation. While implementing this defense may necessitate adjusting the microphone array layout according to its prototype, it effectively raises the bar for adversaries attempting \bd attacks.
Additionally, LipPass~\cite{lu2018lippass} relying on the Doppler effect induced by lip movements during speaking, and the fieldprint from dual-microphone-captured sound field distribution~\cite{yan2019catcher,li2023towards}, are immune to \bd attacks based on voiceprint embedding poisoning.

\color{blue}

\color{black}
\subsection{Limitation and Future Work.} 
In addition to the above discussion of practical deployment, we critically analyze the limitations and future work of \bd as an enrollment-stage attack against speaker recognition systems. The limitations of \bd are that 1) \bd is designed as a framework capable of crafting highly stealthy and effective ultrasonic triggers to contaminate the SRS without being noticed by the legitimate user. Subsequently, the polluted SRS will grant access to both the legitimate user and the adversary with high confidence. However, each trigger is optimized for a specific victim-adversary pair, which may not generalize to grant multiple adversaries to fool the SRS simultaneously. In future work, we envision that crafting full-spectrum triggers, rather than sparse frequencies, can facilitate a more generic enrollment-stage backdoor. 
2) \bd has been proved to successfully attack against SRS in black-box scenarios and can transfer to other unseen SRS models, achieving ACCs and ASRs over 97\%. However, when applied to access control systems that may utilize large-scale SRS models (e.g., WavLM-Xvec) deployed on cloud servers, \bd’s performance tends to decline. We will explore methods to maintain \bd’s high transferability across unseen SRS models, regardless of their model structures or scales.
3) When the SRS continuously updates its speaker database, the effectiveness of \bd may diminish. For instance, certain SRSs may periodically collect voice samples from users to update their stored voiceprints, resulting in a shift of voiceprints towards the victim and away from the adversary. One possible countermeasure is to incorporate the adversary’s trigger in the recognition phase as well. We will investigate this possibility in our future work.

\color{black}

\begin{table*}\centering
\begin{threeparttable}[t]
\setlength{\abovecaptionskip}{0pt}%
\setlength{\belowcaptionskip}{0pt}%
\color{black}
\caption{Comparsion with Prior Works.}

\begin{tabular}{@{}lccccccc@{}}
\toprule
    \textbf{Method} 	& \textbf{Attack Phase} & \textbf{Knowledge$^\#$}	 & \textbf{Task} &  \textbf{Constraint$^\natural$} 				 & \textbf{Audibility$^\downarrow$} & \textbf{Sync.-Free$^\dagger$}  & \textbf{Physical Range$^\ddagger$} \\ \midrule
Zhai \textit{et al.}~\cite{zhai2021backdoor}   & Training-stage &  Black-box \& Data  &  SV & N/A & Monotone & \xmark & N/A \\ \midrule
Koffas~\cite{koffas2021can}   & Training-stage & Black-box \& Data  & SV & N/A & Inaudible & \xmark & N/A   \\ \midrule
PhaseBack~\cite{ye2023stealthy}  & Training-stage & Black-box \& Data & CSI & Phase & Slight Noise & \xmark & N/A \\ \midrule
Abdullah~\cite{abdullah2019ndss}   & Recognition-stage  & Black-box & CSI, SV  & N/A & Intense Noise  & \xmark & 0.3m  \\ \midrule
Zhang \textit{et al.}~\cite{zhang2021attack} & Recognition-stage & White-box & CSI &  $\epsilon$ & Noise & \xmark & 1.7m \\ \midrule
Xie \textit{et al.}~\cite{xie2021enabling} & Recognition-stage & White-box & CSI &  L2-norm, $\epsilon$ & Noise & \xmark & N/A \\ \midrule
Li \textit{et al.}~\cite{li2020practical} & Recognition-stage & White-box & CSI &  L2-norm, $\epsilon$ & Noise & \xmark & 1m \\ \midrule
AdvPulse~\cite{li2020advpulse}   & Recognition-stage & White-box & CSI &  L2-norm, $\epsilon$ & Ambient & \cmark & 2.7m \\ \midrule
FakeBob~\cite{chen2021fakebob}   & Recognition-stage & Black-box & OSI, CSI, SV &  $\epsilon$ & Noise &  \xmark & 2m    \\ \midrule


\textbf{Ours}   & \textbf{Enrollment-stage} &  \textbf{Black-box}   & \textbf{OSI, CSI, SV}  &   \textbf{None}  & \textbf{Inaudible}  & \textbf{\cmark} & \textbf{6m}   
\\ \bottomrule
\end{tabular}

\begin{tablenotes}[flushleft]
\item[] \vspace{-2pt}\hspace{-2pt}\small (i) $^\#$: Knowledge: Data is short for ``access to training data'', which is the inherent strong assumption of backdoor attacks. 
 (ii) $^\natural$: The constraints used to guarantee imperceptibility during optimization. N/A is short for ``Not Applicable'', meaning the method does not consider imperceptibility to humans. $\epsilon$ means limiting the absolute magnitude of perturbations with a constant $\epsilon$. $L_2$-Norm means adding an $L_2$-Norm term in the objective function. None means no stealthiness constraints. 
 (iii) $^\downarrow$: The objective victim's auditory for attacks. Monotone means the signal with a single frequency. Noise wi/wo ``Slight/Intense'' means the degree of being perceived by a victim.
 (iv) $\dagger$: \cmark~means the attack is synchronization-free, while \xmark~is not. 
 (v) $\ddagger$: N/A is short for ``Not Applicable'', meaning the attack is only feasible in the digital domain. 
\end{tablenotes}

\label{tab:compare}
\end{threeparttable}
\end{table*}

\section{Related Works}
\color{black}
In this section, we comprehensively compare \bd with other existing backdoor and adversarial example attacks on speaker recognition systems in Tab.~\ref{tab:compare}. We also discuss prior works regarding to inaudible voice attacks.

\subsection{Training-stage Backdoor Attacks on SRSs}
Backdoor attacks pose a common threat during the training stage of deep neural networks (DNNs). These attacks intend to secretly embed backdoors in DNNs during the training process~\cite{gu2017badnets,li2022backdoor}. Malicious attackers can activate these hidden backdoors using specific triggers to manipulate the output of the targeted DNN model. However, less work has been proposed to implement backdoor attacks on speaker recognition systems. As demonstrated in Tab.~\ref{tab:compare}, three training-stage backdoor attacks enjoy the advantage of being effective against black-box models, but this is under the strong assumption of having access to training data.
Zhai \textit{et al.}~\cite{zhai2021backdoor} present a digital backdoor attack on SV systems by poisoning training samples with a monotone signal, e.g., 1~kHz. 
Instead of audible triggers, Koffas \textit{et al.}~\cite{koffas2021can} inject high-frequency triggers, such as 23~kHz into the training audio files (48~kHz sampling rate). However, due to the communication and storage considerations, most audio formats in VoIP are limited within 16~kHz. 
PhaseBack~\cite{ye2023stealthy} proposes hiding the trigger inside the phase spectrum, which needs to abide by the requirement of minimizing distortion after inverting the frequency domain to the time domain. Nonetheless, we observe there still remains slight electrical noise, rendering it less physically applicable.
SMA~\cite{zheng2023silent} realizes ultrasonic backdoor attacks to alter the recognition result of user speech. Its implementation involves two key steps: (1) modeling the nonlinear distortion of triggers transmitted through the ultrasonic channel; (2) proactively inserting the modeled triggers into training data to ensure a seamless alignment between the poisoning and attack phases.

\subsection{Recognition-stage Adversarial Example Attacks on SRSs}
There has been a growing interest in exploiting adversarial example (AE) attacks against speaker recognition systems at the recognition stage~\cite{li2020practical,chen2021real,abdullah2019ndss,li2020advpulse,zhang2021attack,xie2021enabling}. Typically, malicious attackers generate audio adversarial examples by adding slight yet unnoticeable perturbations to the original input~\cite{carlini2018audio}. As detailed in Tab.~\ref{tab:compare}, all of these attacks function on the recognition stage, and have validated the feasibility of implementing AE attacks on SRSs in both white-box~\cite{zhang2021attack,xie2021enabling,li2020practical,li2020advpulse} and black-box~\cite{abdullah2019ndss,chen2021real} setting. 
Abdullah \textit{et al.}~\cite{abdullah2019ndss} modifies the signal processing-level features of audios, such as MFCC, to launch attacks against ASVs while keeping unintelligible to human beings, yet this attack exhibits intense noise that can easily alert the victim. 
Mainstream AE attacks~\cite{zhang2021attack,xie2021enabling,li2020practical,li2020advpulse,chen2021fakebob} generally integrate the stealthiness constraint into optimization to minimize the possibility of being perceived by the victim and then generate adversarial perturbations against ASVs. 
However, they are susceptible in the real-world scenarios due to subtle perturbations and are effectively only within a near-field range.

Unlike prior works, \bd presents a new type of attack surface, i.e., the enrollment stage rather than the training stage. Thus, we offers a more practical attack in terms of not requiring the access to training data, while still retaining black-box ability. 
We also accomplish the \bd prototype across all speaker recognition tasks, including OSI, CSI, and SV.
In addition, we uncover the potential of applying ultrasound to redefine classical backdoor/AE attacks on audio-interface systems, because it indeed address audibility challenges and stealthiness constraints. Notably, a broader optimization space brought by ultrasound, enables \bd with synchronization-free capability, and we can further extend the attack range only if the emission power is below a threshold~\cite{iijima2018audio}.

\color{black}

\subsection{Inaudible Voice Attacks}
\bd achieves highly imperceptible and real-time backdoor attack by modulating the low-frequency voiceprint signal (also named baseband) on an ultrasonic carrier. The idea of leveraging the inaudible voice attack to complete the specific voiceprint injection is inspired by previous work~\cite{zhang2017dolphinattack,yan2019feasibility}, which exploits the non-linearity loophole of microphones.  The following work has devoted much effect to improving the performance of this attack. Roy \emph{et al.}~\cite{roy2018inaudible} enhance its attack range up to 25~ft. Yan et al.~\cite{yan2020surfingattack} further validate the feasibility of launching inaudible voice attacks across a long distance (e.g., a 30~ft long) via solid materials. Moreover, CapSpeaker~\cite{ji2022capspeaker} enables modulation and injection of ultrasound via the capacitors assembled in smart devices, which provides a new type of attack source for inaudible voice attacks. Vrifle~\cite{li2024vrifle} proposes the first ultrasonic adversarial perturbation attacks that can manipulate ASRs in real time, based on the ultrasound transformation modeling, while addressing user auditory and disruption issues even launching attacks at far distances. 
Nonetheless, existing inaudible voice attacks present poor performance in attacking ASV systems due to severally lossy audio quality after the demodulation of the microphone. Meanwhile, the performances of those attacks are remarkably affected by the placement of attacking equipment. To address the above issues, we carefully craft and optimize the baseband of \bd only on several single-frequency points, making it more robust to attack ASV systems.

\section{Conclusion}

We propose a novel enrollment-stage backdoor attack framework via adversarial ultrasound---\bd, which allows both the legitimate user and the adversary to pass the speaker recognition systems at the recognition stage. By modulating backdoor triggers on the adversarial ultrasound carrier and augmenting the optimization process with multi-factor randomness and robustness, we achieve \bd in a highly imperceptible, universal, and physically robust manner. Moreover, we improve the robustness of adversarial ultrasound in physical world by optimizing the backdoor pattern with only sparse frequencies and adopting the pre-compensation along with single-sideband modulation. \blue{Extensive experiments across various configurations on both digital and physical scenarios demonstrate the effectiveness and robustness of our proposed attack.
To mitigate this newly discovered threat, we also discuss on potential countermeasures, limitations, and future works of this new threat.}

\section*{Acknowledgement}
We thank the anonymous reviewers for their valuable comments and suggestions. This work is supported by China NSFC Grant 62201503, 61925109, 62222114, 62071428, and 62271280. This work is also supported by the Fundamental Research Funds for the Central Universities 226-2022-00223.



 
%

\footnotesize
\bibliographystyle{IEEEtran}
\bibliography{refs}

\begin{thebibliography}{10}
\providecommand{\url}[1]{#1}
\csname url@samestyle\endcsname
\providecommand{\newblock}{\relax}
\providecommand{\bibinfo}[2]{#2}
\providecommand{\BIBentrySTDinterwordspacing}{\spaceskip=0pt\relax}
\providecommand{\BIBentryALTinterwordstretchfactor}{4}
\providecommand{\BIBentryALTinterwordspacing}{\spaceskip=\fontdimen2\font plus
\BIBentryALTinterwordstretchfactor\fontdimen3\font minus
  \fontdimen4\font\relax}
\providecommand{\BIBforeignlanguage}[2]{{%
\expandafter\ifx\csname l@#1\endcsname\relax
\typeout{** WARNING: IEEEtran.bst: No hyphenation pattern has been}%
\typeout{** loaded for the language `#1'. Using the pattern for}%
\typeout{** the default language instead.}%
\else
\language=\csname l@#1\endcsname
\fi
#2}}
\providecommand{\BIBdecl}{\relax}
\BIBdecl

\bibitem{voice_banking}
D.~F. Ben~Gran, ``How banking virtual assistants can improve your banking
  experience,'' Website, 2022,
  \url{https://www.forbes.com/advisor/banking/banking-virtual-assistants/}.

\bibitem{wang2020differences}
S.~Wang, J.~Cao, X.~He, K.~Sun, and Q.~Li, ``When the differences in frequency
  domain are compensated: Understanding and defeating modulated replay attacks
  on automatic speech recognition,'' in \emph{Proceedings of the 2020 ACM
  SIGSAC Conference on Computer and Communications Security}, 2020, pp.
  1103--1119.

\bibitem{blue2018hello}
L.~Blue, L.~Vargas, and P.~Traynor, ``Hello, is it me you're looking for?
  differentiating between human and electronic speakers for voice interface
  security,'' in \emph{Proceedings of the 11th ACM Conference on Security \&
  Privacy in Wireless and Mobile Networks}, 2018, pp. 123--133.

\bibitem{zhai2021backdoor}
T.~Zhai, Y.~Li, Z.~Zhang, B.~Wu, Y.~Jiang, and S.-T. Xia, ``Backdoor attack
  against speaker verification,'' in \emph{ICASSP 2021-2021 IEEE International
  Conference on Acoustics, Speech and Signal Processing (ICASSP)}.\hskip 1em
  plus 0.5em minus 0.4em\relax IEEE, 2021, pp. 2560--2564.

\bibitem{gu2017badnets}
T.~Gu, B.~Dolan-Gavitt, and S.~Garg, ``Badnets: Identifying vulnerabilities in
  the machine learning model supply chain,'' \emph{arXiv preprint
  arXiv:1708.06733}, vol.~1, 2017.

\bibitem{shi2022audio}
C.~Shi, T.~Zhang, Z.~Li, H.~Phan, T.~Zhao, Y.~Wang, J.~Liu, B.~Yuan, and
  Y.~Chen, ``Audio-domain position-independent backdoor attack via unnoticeable
  triggers,'' in \emph{Proceedings of the 28th Annual International Conference
  on Mobile Computing And Networking}, 2022, pp. 583--595.

\bibitem{luo2022practical}
Y.~Luo, J.~Tai, X.~Jia, and S.~Zhang, ``Practical backdoor attack against
  speaker recognition system,'' in \emph{Information Security Practice and
  Experience: 17th International Conference, ISPEC 2022, Taipei, Taiwan,
  November 23--25, 2022, Proceedings}.\hskip 1em plus 0.5em minus 0.4em\relax
  Springer, 2022, pp. 468--484.

\bibitem{chen2021real}
G.~Chen, S.~Chenb, L.~Fan, X.~Du, Z.~Zhao, F.~Song, and Y.~Liu, ``Who is real
  bob? adversarial attacks on speaker recognition systems,'' in \emph{2021 IEEE
  Symposium on Security and Privacy (SP)}.\hskip 1em plus 0.5em minus
  0.4em\relax IEEE, 2021, pp. 694--711.

\bibitem{li2020advpulse}
Z.~Li, Y.~Wu, J.~Liu, Y.~Chen, and B.~Yuan, ``Advpulse: Universal,
  synchronization-free, and targeted audio adversarial attacks via subsecond
  perturbations,'' in \emph{Proceedings of the 2020 ACM SIGSAC Conference on
  Computer and Communications Security}, 2020, pp. 1121--1134.

\bibitem{li2020practical}
Z.~Li, C.~Shi, Y.~Xie, J.~Liu, B.~Yuan, and Y.~Chen, ``Practical adversarial
  attacks against speaker recognition systems,'' in \emph{Proceedings of the
  21st international workshop on mobile computing systems and applications},
  2020, pp. 9--14.

\bibitem{zhang2017dolphinattack}
G.~Zhang, C.~Yan, X.~Ji, T.~Zhang, T.~Zhang, and W.~Xu, ``Dolphinattack:
  Inaudible voice commands,'' in \emph{Proceedings of the 2017 ACM SIGSAC
  conference on computer and communications security}, 2017, pp. 103--117.

\bibitem{lee2006efficient}
H.~Lee, A.~Battle, R.~Raina, and A.~Ng, ``Efficient sparse coding algorithms,''
  \emph{Advances in neural information processing systems}, vol.~19, 2006.

\bibitem{zeng2021rethinking}
Y.~Zeng, W.~Park, Z.~M. Mao, and R.~Jia, ``Rethinking the backdoor attacks'
  triggers: A frequency perspective,'' in \emph{Proceedings of the IEEE/CVF
  International Conference on Computer Vision}, 2021, pp. 16\,473--16\,481.

\bibitem{wierstra2014natural}
D.~Wierstra, T.~Schaul, T.~Glasmachers, Y.~Sun, J.~Peters, and J.~Schmidhuber,
  ``Natural evolution strategies,'' \emph{The Journal of Machine Learning
  Research}, vol.~15, no.~1, pp. 949--980, 2014.

\bibitem{2020ecapatdnn}
B.~Desplanques, J.~Thienpondt, and K.~Demuynck, ``{ECAPA-TDNN:} emphasized
  channel attention, propagation and aggregation in {TDNN} based speaker
  verification,'' in \emph{Interspeech 2020, 21st Annual Conference of the
  International Speech Communication Association}, 2020, pp. 3830--3834.

\bibitem{bredin2020pyannote}
H.~Bredin, R.~Yin, J.~M. Coria, G.~Gelly, P.~Korshunov, M.~Lavechin, D.~Fustes,
  H.~Titeux, W.~Bouaziz, and M.-P. Gill, ``Pyannote. audio: neural building
  blocks for speaker diarization,'' in \emph{ICASSP 2020-2020 IEEE
  International Conference on Acoustics, Speech and Signal Processing
  (ICASSP)}.\hskip 1em plus 0.5em minus 0.4em\relax IEEE, 2020, pp. 7124--7128.

\bibitem{xie2019utterance}
W.~Xie, A.~Nagrani, J.~S. Chung, and A.~Zisserman, ``Utterance-level
  aggregation for speaker recognition in the wild,'' in \emph{ICASSP 2019-2019
  IEEE International Conference on Acoustics, Speech and Signal Processing
  (ICASSP)}.\hskip 1em plus 0.5em minus 0.4em\relax IEEE, 2019, pp. 5791--5795.

\bibitem{chen2022wavlm}
S.~Chen, C.~Wang, Z.~Chen, Y.~Wu, S.~Liu, Z.~Chen, J.~Li, N.~Kanda,
  T.~Yoshioka, X.~Xiao \emph{et~al.}, ``Wavlm: Large-scale self-supervised
  pre-training for full stack speech processing,'' \emph{IEEE Journal of
  Selected Topics in Signal Processing}, vol.~16, no.~6, pp. 1505--1518, 2022.

\bibitem{koluguri2020speakernet}
N.~R. Koluguri, J.~Li, V.~Lavrukhin, and B.~Ginsburg, ``Speakernet: 1d
  depth-wise separable convolutional network for text-independent speaker
  recognition and verification,'' \emph{arXiv preprint arXiv:2010.12653}, 2020.

\bibitem{wan2018generalized}
L.~Wan, Q.~Wang, A.~Papir, and I.~L. Moreno, ``Generalized end-to-end loss for
  speaker verification,'' in \emph{2018 IEEE International Conference on
  Acoustics, Speech and Signal Processing (ICASSP)}.\hskip 1em plus 0.5em minus
  0.4em\relax IEEE, 2018, pp. 4879--4883.

\bibitem{chung2020in}
J.~S. Chung, J.~Huh, S.~Mun, M.~Lee, H.~S. Heo, S.~Choe, C.~Ham, S.~Jung, B.-J.
  Lee, and I.~Han, ``In defence of metric learning for speaker recognition,''
  in \emph{Interspeech}, 2020.

\bibitem{nagrani2017voxceleb}
A.~Nagrani, J.~S. Chung, and A.~Zisserman, ``Voxceleb: a large-scale speaker
  identification dataset,'' \emph{arXiv preprint arXiv:1706.08612}, vol.~1,
  2017.

\bibitem{panayotov2015librispeech}
V.~Panayotov, G.~Chen, D.~Povey, and S.~Khudanpur, ``Librispeech: an asr corpus
  based on public domain audio books,'' in \emph{2015 IEEE international
  conference on acoustics, speech and signal processing (ICASSP)}.\hskip 1em
  plus 0.5em minus 0.4em\relax IEEE, 2015, pp. 5206--5210.

\bibitem{jajodia2011encyclopedia}
S.~Jajodia and H.~C. van~van Tilborg, \emph{Encyclopedia of Cryptography and
  Security: L-Z}.\hskip 1em plus 0.5em minus 0.4em\relax Springer, 2011.

\bibitem{sood2021speech}
M.~Sood and S.~Jain, ``Speech recognition employing mfcc and dynamic time
  warping algorithm,'' in \emph{Innovations in Information and Communication
  Technologies (IICT-2020)}.\hskip 1em plus 0.5em minus 0.4em\relax Springer,
  2021, pp. 235--242.

\bibitem{thian2004spectral}
N.~P.~H. Thian, C.~Sanderson, and S.~Bengio, ``Spectral subband centroids as
  complementary features for speaker authentication,'' in \emph{International
  conference on biometric authentication}.\hskip 1em plus 0.5em minus
  0.4em\relax Springer, 2004, pp. 631--639.

\bibitem{hermansky1990perceptual}
H.~Hermansky, ``Perceptual linear predictive (plp) analysis of speech,''
  \emph{the Journal of the Acoustical Society of America}, vol.~87, no.~4, pp.
  1738--1752, 1990.

\bibitem{dehak2010front}
N.~Dehak, P.~J. Kenny, R.~Dehak, P.~Dumouchel, and P.~Ouellet, ``Front-end
  factor analysis for speaker verification,'' \emph{IEEE Transactions on Audio,
  Speech, and Language Processing}, vol.~19, no.~4, pp. 788--798, 2010.

\bibitem{snyder2018x}
D.~Snyder, D.~Garcia-Romero, G.~Sell, D.~Povey, and S.~Khudanpur, ``X-vectors:
  Robust dnn embeddings for speaker recognition,'' in \emph{2018 IEEE
  international conference on acoustics, speech and signal processing
  (ICASSP)}.\hskip 1em plus 0.5em minus 0.4em\relax IEEE, 2018, pp. 5329--5333.

\bibitem{he2022ok}
R.~He, X.~Ji, X.~Li, Y.~Cheng, and W.~Xu, ``“ok, siri” or “hey,
  google”: Evaluating voiceprint distinctiveness via content-based prole
  score,'' in \emph{Proceedings of the 31th USENIX Security Symposium}, 2022.

\bibitem{Furui2008SRS}
S.~Furui, ``Speaker recognition,''
  \url{http://www.scholarpedia.org/article/Speaker_recognition}, 2008.

\bibitem{roy2018inaudible}
N.~Roy, S.~Shen, H.~Hassanieh, and R.~R. Choudhury, ``Inaudible voice commands:
  The long-range attack and defense,'' in \emph{Proceedings of the 15th USENIX
  Symposium on Networked Systems Design and Implementation (NSDI 18)}, 2018,
  pp. 547--560.

\bibitem{ji2022capspeaker}
X.~Ji, J.~Zhang, S.~Jiang, J.~Li, and W.~Xu, ``Capspeaker: Injecting voices to
  microphones via capacitors,'' in \emph{Proceedings of the 2021 ACM SIGSAC
  Conference on Computer and Communications Security}, 2021.

\bibitem{yan2020surfingattack}
Q.~Yan, K.~Liu, Q.~Zhou, H.~Guo, and N.~Zhang, ``Surfingattack: Interactive
  hidden attack on voice assistants using ultrasonic guided waves,'' in
  \emph{Proceedings of the Network and Distributed Systems Security (NDSS)
  Symposium}, 2020.

\bibitem{sugawara2020light}
T.~Sugawara, B.~Cyr, S.~Rampazzi, D.~Genkin, and K.~Fu, ``Light commands:
  laser-based audio injection attacks on voice-controllable systems,'' in
  \emph{Proceedings of the 29th USENIX Security Symposium (USENIX Security
  20)}, 2020, pp. 2631--2648.

\bibitem{wang2022ghosttalk}
Y.~Wang, H.~Guo, and Q.~Yan, ``Ghosttalk: Interactive attack on smartphone
  voice system through power line,'' in \emph{Network and Distributed Systems
  Security (NDSS) Symposium}, 2022.

\bibitem{ai2019improvement}
H.~Ai, Y.~Wang, Y.~Yang, and Q.~Zhang, ``An improvement of the degradation of
  speaker recognition in continuous cold speech for home assistant,'' in
  \emph{International Symposium on Cyberspace Safety and Security}.\hskip 1em
  plus 0.5em minus 0.4em\relax Springer, 2019, pp. 363--373.

\bibitem{o2022evaluating}
B.~O'Brien, C.~Meunier, and A.~Ghio, ``Evaluating the effects of modified
  speech on perceptual speaker identification performance,'' in
  \emph{Interspeech 2022}, 2022.

\bibitem{tull1996analysis}
R.~G. Tull and J.~C. Rutledge, ``Analysis of ``cold-affected'' speech for
  inclusion in speaker recognition systems.'' \emph{The Journal of the
  Acoustical Society of America}, vol.~99, no.~4, pp. 2549--2574, 1996.

\bibitem{wagner2017infected}
\BIBentryALTinterwordspacing
J.~Wagner, T.~Fraga{-}Silva, Y.~Josse, D.~Schiller, A.~Seiderer, and
  E.~Andr{\'{e}}, ``Infected phonemes: How a cold impairs speech on a phonetic
  level,'' in \emph{Interspeech 2017, 18th Annual Conference of the
  International Speech Communication Association, Stockholm, Sweden, August
  20-24, 2017}, F.~Lacerda, Ed.\hskip 1em plus 0.5em minus 0.4em\relax {ISCA},
  2017, pp. 3457--3461. [Online]. Available:
  \url{http://www.isca-speech.org/archive/Interspeech\_2017/abstracts/1066.html}
\BIBentrySTDinterwordspacing

\bibitem{zheng2020automatic}
L.~Zheng, J.~Li, M.~Sun, X.~Zhang, and T.~F. Zheng, ``When automatic voice
  disguise meets automatic speaker verification,'' \emph{IEEE Transactions on
  Information Forensics and Security}, vol.~16, pp. 824--837, 2020.

\bibitem{tavi2022improving}
L.~Tavi, T.~Kinnunen, and R.~G. Hautam{\"a}ki, ``Improving speaker
  de-identification with functional data analysis of f0 trajectories,''
  \emph{Speech Communication}, vol. 140, pp. 1--10, 2022.

\bibitem{manocha2020differentiable}
P.~Manocha, A.~Finkelstein, R.~Zhang, N.~J. Bryan, G.~J. Mysore, and Z.~Jin,
  ``A differentiable perceptual audio metric learned from just noticeable
  differences,'' \emph{arXiv preprint arXiv:2001.04460}, 2020.

\bibitem{tibshirani1996regression}
R.~Tibshirani, ``Regression shrinkage and selection via the lasso,''
  \emph{Journal of the Royal Statistical Society: Series B (Methodological)},
  vol.~58, no.~1, pp. 267--288, 1996.

\bibitem{Sound}
R.~E. Berg, ``{Sound Physics},''
  \url{https://www.britannica.com/science/sound-physics}, 2019.

\bibitem{zhang2021eararray}
G.~Zhang, X.~Ji, X.~Li, G.~Qu, and W.~Xu, ``Eararray: Defending against
  dolphinattack via acoustic attenuation,'' in \emph{Network and Distributed
  System Security (NDSS) Symposium}, 2021.

\bibitem{jeub2009binaural}
M.~Jeub, M.~Schafer, and P.~Vary, ``A binaural room impulse response database
  for the evaluation of dereverberation algorithms,'' in \emph{2009 16th
  International Conference on Digital Signal Processing}.\hskip 1em plus 0.5em
  minus 0.4em\relax IEEE, 2009, pp. 1--5.

\bibitem{deng2022fencesitter}
J.~Deng, Y.~Chen, and W.~Xu, ``Fencesitter: Black-box, content-agnostic, and
  synchronization-free enrollment-phase attacks on speaker recognition
  systems,'' in \emph{Proceedings of the 2021 ACM SIGSAC Conference on Computer
  and Communications Security}, 2022.

\bibitem{li2023learning}
X.~Li, X.~Ji, C.~Yan, C.~Li, Y.~Li, Z.~Zhang, and W.~Xu, ``Learning normality
  is enough: A software-based mitigation against the inaudible voice attacks,''
  in \emph{Proceedings of the 32nd USENIX Security Symposium}, 2023.

\bibitem{infomasker2023}
P.~Huang, Y.~Wei, P.~Cheng, Z.~Ba, L.~Lu, F.~Lin, F.~Zhang, and K.~Ren,
  ``Infomasker: Preventing eavesdropping using phoneme-based noise,'' in
  \emph{Network and Distributed System Security (NDSS) Symposium}, 2023.

\bibitem{kingma2014adam}
D.~P. Kingma and J.~Ba, ``Adam: A method for stochastic optimization,''
  \emph{arXiv preprint arXiv:1412.6980}, vol.~1, 2014.

\bibitem{Keysight}
K.~Technologies, ``{EXG X-Series Signal Generator},''
  \url{https://www.keysight.com/us/en/assets/7018-03381/data-sheets/5991-0039.pdf},
  2019.

\bibitem{amplifierHSA4015}
Micronix, ``Nf hsa4015,'' Website, 2013,
  \url{https://eshop.micronix.eu/measurement-equipment/electrical-quantities/nf-corporation-instruments/high-speed-bipolar-amplifiers/hsa-4051.html}.

\bibitem{aishell_2017}
H.~Bu, J.~Du, X.~Na, B.~Wu, and H.~Zheng, ``Aishell-1: An open-source mandarin
  speech corpus and a speech recognition baseline,'' in \emph{Proceedings of
  the Oriental COCOSDA 2017}, 2017, p. Submitted.

\bibitem{Pratap2020MLSAL}
V.~Pratap, Q.~Xu, A.~Sriram, G.~Synnaeve, and R.~Collobert, ``Mls: A
  large-scale multilingual dataset for speech research,'' \emph{ArXiv}, vol.
  abs/2012.03411, 2020.

\bibitem{zeinali2019short}
H.~Zeinali, K.~A. Lee, J.~Alam, and L.~Burget, ``Short-duration speaker
  verification (sdsv) challenge 2021: the challenge evaluation plan,''
  \emph{arXiv preprint arXiv:1912.06311}, vol.~1, 2019.

\bibitem{thienpondt2020cross}
J.~Thienpondt, B.~Desplanques, and K.~Demuynck, ``Cross-lingual speaker
  verification with domain-balanced hard prototype mining and
  language-dependent score normalization,'' \emph{arXiv preprint
  arXiv:2007.07689}, 2020.

\bibitem{jin2007far}
Q.~Jin, T.~Schultz, and A.~Waibel, ``Far-field speaker recognition,''
  \emph{IEEE Transactions on Audio, Speech, and Language Processing}, vol.~15,
  no.~7, pp. 2023--2032, 2007.

\bibitem{gusev2020deep}
A.~Gusev, V.~Volokhov, T.~Andzhukaev, S.~Novoselov, G.~Lavrentyeva, M.~Volkova,
  A.~Gazizullina, A.~Shulipa, A.~Gorlanov, A.~Avdeeva \emph{et~al.}, ``Deep
  speaker embeddings for far-field speaker recognition on short utterances,''
  \emph{arXiv preprint arXiv:2002.06033}, 2020.

\bibitem{iijima2018audio}
R.~Iijima, S.~Minami, Z.~Yunao, T.~Takehisa, T.~Takahashi, Y.~Oikawa, and
  T.~Mori, ``Audio hotspot attack: An attack on voice assistance systems using
  directional sound beams,'' in \emph{Proceedings of the 2018 ACM SIGSAC
  Conference on Computer and Communications Security}, 2018, pp. 2222--2224.

\bibitem{mandasari2012effect}
M.~I. Mandasari, M.~McLaren, and D.~A. van Leeuwen, ``The effect of noise on
  modern automatic speaker recognition systems,'' in \emph{2012 IEEE
  International Conference on Acoustics, Speech and Signal Processing
  (ICASSP)}.\hskip 1em plus 0.5em minus 0.4em\relax IEEE, 2012, pp. 4249--4252.

\bibitem{freesound}
freesound, ``freesound.org,'' \url{https://freesound.org/}, 2022.

\bibitem{yuan2018commandersong}
X.~Yuan, Y.~Chen, Y.~Zhao, Y.~Long, X.~Liu, K.~Chen, S.~Zhang, H.~Huang,
  X.~Wang, and C.~A. Gunter, ``Commandersong: {A} systematic approach for
  practical adversarial voice recognition,'' in \emph{27th {USENIX} Security
  Symposium, {USENIX} Security}, 2018, pp. 49--64.

\bibitem{li2024vrifle}
X.~Li, C.~Yan, X.~Lu, X.~Ji, and W.~Xu, ``Inaudible adversarial perturbation:
  Manipulating the recognition of user speech in real time,'' in \emph{Network
  and Distributed System Security (NDSS) Symposium}, 2024.

\bibitem{he2019canceling}
Y.~He, J.~Bian, X.~Tong, Z.~Qian, W.~Zhu, X.~Tian, and X.~Wang, ``Canceling
  inaudible voice commands against voice control systems,'' in
  \emph{Proceedings of the 25th Annual International Conference on Mobile
  Computing and Networking}, 2019, pp. 1--15.

\bibitem{lu2018lippass}
L.~Lu, J.~Yu, Y.~Chen, H.~Liu, Y.~Zhu, Y.~Liu, and M.~Li, ``Lippass: Lip
  reading-based user authentication on smartphones leveraging acoustic
  signals,'' in \emph{IEEE INFOCOM 2018-IEEE Conference on Computer
  Communications}.\hskip 1em plus 0.5em minus 0.4em\relax IEEE, 2018, pp.
  1466--1474.

\bibitem{yan2019catcher}
C.~Yan, Y.~Long, X.~Ji, and W.~Xu, ``The catcher in the field: A fieldprint
  based spoofing detection for text-independent speaker verification,'' in
  \emph{Proceedings of the 2019 ACM SIGSAC Conference on Computer and
  Communications Security}, 2019, pp. 1215--1229.

\bibitem{li2023towards}
X.~Li, Z.~Zheng, C.~Yan, C.~Li, X.~Ji, and W.~Xu, ``Towards pitch-insensitive
  speaker verification via soundfield,'' \emph{IEEE Internet of Things
  Journal}, 2023.

\bibitem{koffas2021can}
S.~Koffas, J.~Xu, M.~Conti, and S.~Picek, ``Can you hear it? backdoor attacks
  via ultrasonic triggers,'' \emph{arXiv preprint arXiv:2107.14569}, vol.~1,
  2021.

\bibitem{ye2023stealthy}
Z.~Ye, D.~Yan, L.~Dong, J.~Deng, and S.~Yu, ``Stealthy backdoor attack against
  speaker recognition using phase-injection hidden trigger,'' \emph{IEEE Signal
  Processing Letters}, 2023.

\bibitem{abdullah2019ndss}
H.~Abdullah, W.~Garcia, C.~Peeters, P.~Traynor, K.~R.~B. Butler, and J.~Wilson,
  ``Practical hidden voice attacks against speech and speaker recognition
  systems,'' in \emph{26th Annual Network and Distributed System Security
  Symposium, {NDSS} 2019}, 2019.

\bibitem{zhang2021attack}
W.~Zhang, S.~Zhao, L.~Liu, J.~Li, X.~Cheng, T.~F. Zheng, and X.~Hu, ``Attack on
  practical speaker verification system using universal adversarial
  perturbations,'' in \emph{ICASSP 2021-2021 IEEE International Conference on
  Acoustics, Speech and Signal Processing (ICASSP)}.\hskip 1em plus 0.5em minus
  0.4em\relax IEEE, 2021, pp. 2575--2579.

\bibitem{xie2021enabling}
Y.~Xie, Z.~Li, C.~Shi, J.~Liu, Y.~Chen, and B.~Yuan, ``Enabling fast and
  universal audio adversarial attack using generative model,'' in
  \emph{Proceedings of the AAAI Conference on Artificial Intelligence},
  vol.~35, 2021, pp. 14\,129--14\,137.

\bibitem{chen2021fakebob}
G.~Chen, S.~Chen, L.~Fan, X.~Du, Z.~Zhao, F.~Song, and Y.~Liu, ``Who is real
  bob? adversarial attacks on speaker recognition systems,'' in \emph{42nd
  {IEEE} Symposium on Security and Privacy, {SP} 2021}.\hskip 1em plus 0.5em
  minus 0.4em\relax {IEEE}, 2021, pp. 694--711.

\bibitem{li2022backdoor}
Y.~Li, Y.~Jiang, Z.~Li, and S.-T. Xia, ``Backdoor learning: A survey,''
  \emph{IEEE Transactions on Neural Networks and Learning Systems}, vol.~1,
  2022.

\bibitem{zheng2023silent}
Z.~Zheng, X.~Li, C.~Yan, X.~Ji, and W.~Xu, ``The silent manipulator: A
  practical and inaudible backdoor attack against speech recognition systems,''
  in \emph{Proceedings of the 31st ACM International Conference on Multimedia},
  2023, pp. 7849--7858.

\bibitem{carlini2018audio}
N.~Carlini and D.~Wagner, ``Audio adversarial examples: Targeted attacks on
  speech-to-text,'' in \emph{2018 IEEE security and privacy workshops
  (SPW)}.\hskip 1em plus 0.5em minus 0.4em\relax IEEE, 2018, pp. 1--7.

\bibitem{yan2019feasibility}
C.~Yan, G.~Zhang, X.~Ji, T.~Zhang, T.~Zhang, and W.~Xu, ``The feasibility of
  injecting inaudible voice commands to voice assistants,'' \emph{IEEE
  Transactions on Dependable and Secure Computing}, vol.~18, no.~3, pp.
  1108--1124, 2019.

\end{thebibliography}


 





\end{document}